\documentclass[aps,twocolumn,preprintnumbers,nofootinbib,superscriptaddress]{revtex4-1}
\usepackage{eurosym}
\usepackage{amsmath}
\usepackage{graphicx}
\usepackage{amsfonts}
\usepackage{array}
\usepackage{amsthm}
\usepackage{bm}
\usepackage{supertabular}
\usepackage{subfig}
\usepackage[breaklinks]{hyperref}
\usepackage{graphicx}
\usepackage{caption}
\usepackage[font=small,labelfont=bf,justification=raggedright]{caption}
\usepackage{color}
\usepackage{booktabs}
\usepackage{multirow}
\newcommand{\be}{\begin{equation}}
	\newcommand{\ee}{\end{equation}}
\newcommand{\ba}{\begin{eqnarray}}
	\newcommand{\ea}{\end{eqnarray}}
\newcommand{\bal}{\begin{align}}
	\newcommand{\eal}{\end{align}}
\newcommand{\lb}{\label}

\newcommand{\bw}{\begin{widetext}}
	\newcommand{\ew}{\end{widetext}}

\usepackage[utf8]{inputenc}
\usepackage{amssymb}
\usepackage{amsmath}
\usepackage{amsfonts}
\usepackage{graphicx, float}
\usepackage{graphicx, epsfig}
\usepackage{color}
\usepackage{enumerate}
\newcommand{\beq}{\begin{equation}}
	\newcommand{\eeq}{\end{equation}}
\newcommand{\bqn}{\begin{eqnarray}}
	\newcommand{\eqn}{\end{eqnarray}}

\begin{document}
%	\title{Relating Shadow Radius to Quasinormal Modes and Quasiperiodic Oscillations in Spacetime of Quantum Deformed Kerr Black Holes}
%Relating shadow to quasinormal modes, thermodynamical stability, and
%    quasiperiodic oscillations of quantum deformed Kerr black holes
		\title{Equatorial and polar quasinormal modes and quasiperiodic oscillations of quantum deformed Kerr black hole}
	%\title{On the Interrelationship of Shadow Radius and Quasinormal Modes to Keplerian for a Quantum Deformed Kerr Black Hole}
	%Shadows of Rotating Quantum Deformed BHs}
	%\title{Shadow and deflection angle of a rotating regular BH in conformal massive gravity}
	%\title{Rotating regular BHs in conformal massive gravity and their shadows}
	\author{Kimet Jusufi}
	\email{kimet.jusufi@unite.edu.mk (corresponding author)}
	\affiliation{Physics Department, State University of Tetovo, Ilinden Street nn, 1200, Tetovo, North Macedonia}
	\author{Mustapha Azreg-A\"{\i}nou}%\email{?????}
	\affiliation{Ba\c{s}kent University, Engineering Faculty, Ba\u{g}l\i ca Campus, 06790-Ankara, Turkey}

	\author{Mubasher Jamil}
	\affiliation{Institute for Theoretical Physics and Cosmology, Zhejiang University of Technology,
		Hangzhou, 310023 China}
	\affiliation{School of Natural Sciences (SNS), National University of Sciences and Technology (NUST), Islamabad 44000, Pakistan}

	\author{Qiang Wu}
\affiliation{Institute for Theoretical Physics and Cosmology, Zhejiang University of Technology,
Hangzhou, 310023 China}	
	
	\begin{abstract}
		In this paper we focus on the relation between quasinormal modes (QNMs) and a rotating black hole shadows. As a specific example, we consider the quantum deformed Kerr black hole obtained via Newman--Janis--Azreg-A\"{\i}nou  algorithm. In particular, using the geometric-optics correspondence between the parameters of a QNMs and the conserved quantities along geodesics, we show that, in the eikonal limit, the real part of QNMs is related to the Keplerian frequency for equatorial orbits.  To this end, we explore the typical shadow radius for the viewing angles, $\theta_0=\pi/2$, and obtained an interesting relation in the case of viewing angle $\theta_0=0$ (or equivalently $\theta_0=\pi$).  Furthermore we have computed the corresponding equatorial and polar modes and the thermodynamical stability of the quantum deformed Kerr black hole.  We also investigate other astrophysical applications such as the quasiperiodic oscillations and the motion of S2 star to constrain the quantum deforming parameter. 
	\end{abstract}
	
	\maketitle
	\section{Introduction} 
	
	Astrophysical candidates of black holes (BHs) and their immediate environments are quite significant not only to understand high energy astrophysics itself but also to test and constrain various phenomenological models of modified and quantum gravity theories. Most astrophysical BHs are surrounded by various kind of matter distributions such as accretion disks, electromagnetic fields, jets, galactic plasma and dark matter distributions, which can have drastic implications on the stability, evolution, and observations of BHs. The supermassive BH in the center of giant elliptical M87 galaxy interestingly possesses all the above mentioned characteristics and has been intensely investigated both theoretically and observationally. More recently, the detection of shadow cast by the event horizon of M87 back hole has attracted huge attention \cite{m87}, whereas its other empirical features such as magnetic field strength, polarization and mass accretion rate \cite{m871} are of monumental relevance for testing phenomenological models of gravity. The black hole physics recently attracted considerable interest, from the well known Kerr BH in general relativity to BH solutions in other theories such as loop quantum gravity have been studied by testing the rotational nature of these black holes and possible deviations from general relativity as well as the possibility to test fundamental physics, including extra dimensions using the black hole shadow (see \cite{papers} and references therein).
	
	Notably, the discoveries of gravitational waves due to collisions and mergers of intermediate mass BHs by the LIGO/Virgo collaborations have provided an alternative method to constrain phenomenological models \cite{ligo}. In particular, the theoretically developed profiles of BH QNMs can be matched with the observationally obtained signals during the inspiral, merger, and ringdown phases \cite{K1}. In Ref. \cite{cardoso} the null unstable geodesics were  related to the QNMs of BH as well as the Lyapunov exponent which is linked to the instability timescale, while in Ref. \cite{stefanov} a connection between the strong lensing and the QNMs was shown. Recently, Jusufi has suggested that the typical shadow radius of a BH is linked inversely with the real part of the QNM frequency if the eikonal limit is applied in the axisymmetric spacetimes \cite{J1}. More recently, Yang used the geometric correspondence \cite{Yang:2012he} and argued that the shadow seen by an distant observer at a given inclination angle can be mapped to a family of QNMs, thus suggesting the possibility of testing this correspondence with space borne gravitational wave detectors and the next-generation Event Horizon Telescope \cite{Y}. We here pursue a similar goal for the spinning quantum deformed BH.

	Quasiperiodic oscillations or QPOs are high energy astrophysical phenomenon usually associated with the microquasars which are BH or neutron star accretion disk systems or compact BH- neutron star binaries with accretion disks.  The QPOs are usually characterized by low frequency (of order tens Hz) and high frequency (of order hundred to kilo Hz) in the power density X-ray spectrum of the sources. Theoretically, QPOs are studied using a toy model of Kerr BH in the frameworks of Keplerian or forced resonance, relativistic precession or diskoseismology models \cite{B}, while we are mainly interested in the high frequency ones. We would like to mention that so far the exact origin of the QPOs in the microquasars is not known. Generally speaking, they may be some oscillations of the accretion disk. In some phenomenological models of QPOs, the oscillations of the disk have the same frequencies as the oscillations of a particle moving around the BH, so we can simplify the calculations and directly calculate the epicyclic frequencies of oscillating particles. From the analytical and numerical perspective, the QPOs are investigated via the epicyclic (or quasi-cyclic) motion of charged particles around spinning BHs in the presence of magnetic fields \cite{A11,T22} and sometimes using only neutral particles in a curved background without the use of magnetic field. The latter is assumed to be weak such that the spacetime curvature is undisturbed due to the magnetic fields. The charged particles are assumed to move along nearly circular orbits with small perturbations in their orbits. Since the orbit is perturbed negligibly, the motion of charged particles is still described by the perturbed geodesic equations containing perturbed position variables and perturbed four velocities. If the perturbations are small, one can retain only linear terms in all perturbed variables in the governing equations. Thus the governing equations to study epicyclic motion of charged particles are the perturbed geodesic equations and the normalization equation. The latter is employed to decouple certain velocity variables during calculations. By simplifying the above equations, one gets only two independent equations of motion corresponding to the perturbations in the radial and angular position components. The two equations also involve the epicyclic frequency parameters corresponding to each radial and angular perturbations. It turns out that both epicyclic frequencies are determined by not only geometrical but also electromagnetic configurations. In the end, one can relate the analytical radial and angular epicyclic frequencies expressions with the observed upper and lower frequencies respectively, which are obtained from the observations of few galactic microquasars 
	%(for details, one can see \cite{G,A2}). 
	The upper and lower frequencies are then plotted against the radial coordinate by assuming the resonance frequency ratio (usually one takes 3/2 and more generally $n/2$, where $n$ is a Natural number) and the mass error bands. If the frequency curves pass through the mass error bands, one can then constrain the free parameters of the theory. We shall employ the data of upper and lower frequencies of the three microquasars namely, GRO J1655-40, XTEJ1550-564 and GRS 1915+105 \cite{T,R}.

	Various theories of modified gravity and candidates of quantum gravity often introduce additional or correction terms in the exact BH solutions of general relativity. In particular, Kazakov and Solodukhin (KS) investigated a semi-classical gravitational theory which was string theory inspired and renormalizable \cite{def}. Within that framework, they studied general relativity dimensionally reduced to a two dimensional dilaton gravity. An interesting implication of their work is that the curvature singularity ($r=0$) of the Schwarzschild BH is abated due to quantum fluctuations, to the value $r=r_\text{pl}$ which is the Planck length. The singularity attains a geometrical extended structure of a 2-sphere with a finite volume. Note that the new KS solution depends on a particular choice of the dilaton field potential $U$, which if set to unity gives rise to vacuum Einstein field equations. Since than, the KS BH has attracted renewed interest intermittently. Several classical and astrophysical tests of KS theory have been performed using the KS BH as a candidate for an astrophysical BH, see \cite{KS1}. We aim to derive an effective spinning KS BH solution by using the Newman-Janis-Azreg-A\"{i}nou algorithm and investigate its phenomenology using shadows, QNMs, QPOs and stellar dynamics.
	
	The outline of this paper is as follows: In Sec.~\ref{secrqd}, we derive an effective solution of a Kerr BH with quantum deformation using a seed KS solution. In Sec.~\ref{secqnm}, we study QNMs and shadows and discuss their interrelation. In Sec. IV, we investigate the thermodynamical stability in terms of shadow radius and the real part of QNMs. In Sec. V we will study the motion of S2 star to constrain the quantum deformation parameter. In Sec.~\ref{secqpos} we provide another physical mean to restrict the values of the quantum-deforming parameter by seeking curve fitting to the observed values of the QPOs of three microquasars. We present a conclusion in Sec. ~\ref{conc}. The metric signature adopted as $(-,+,+,+)$ and chosen units are $c=G=\hbar=1$.

	\begin{figure*}[ht!]
		\centering
		\includegraphics[scale=0.62]{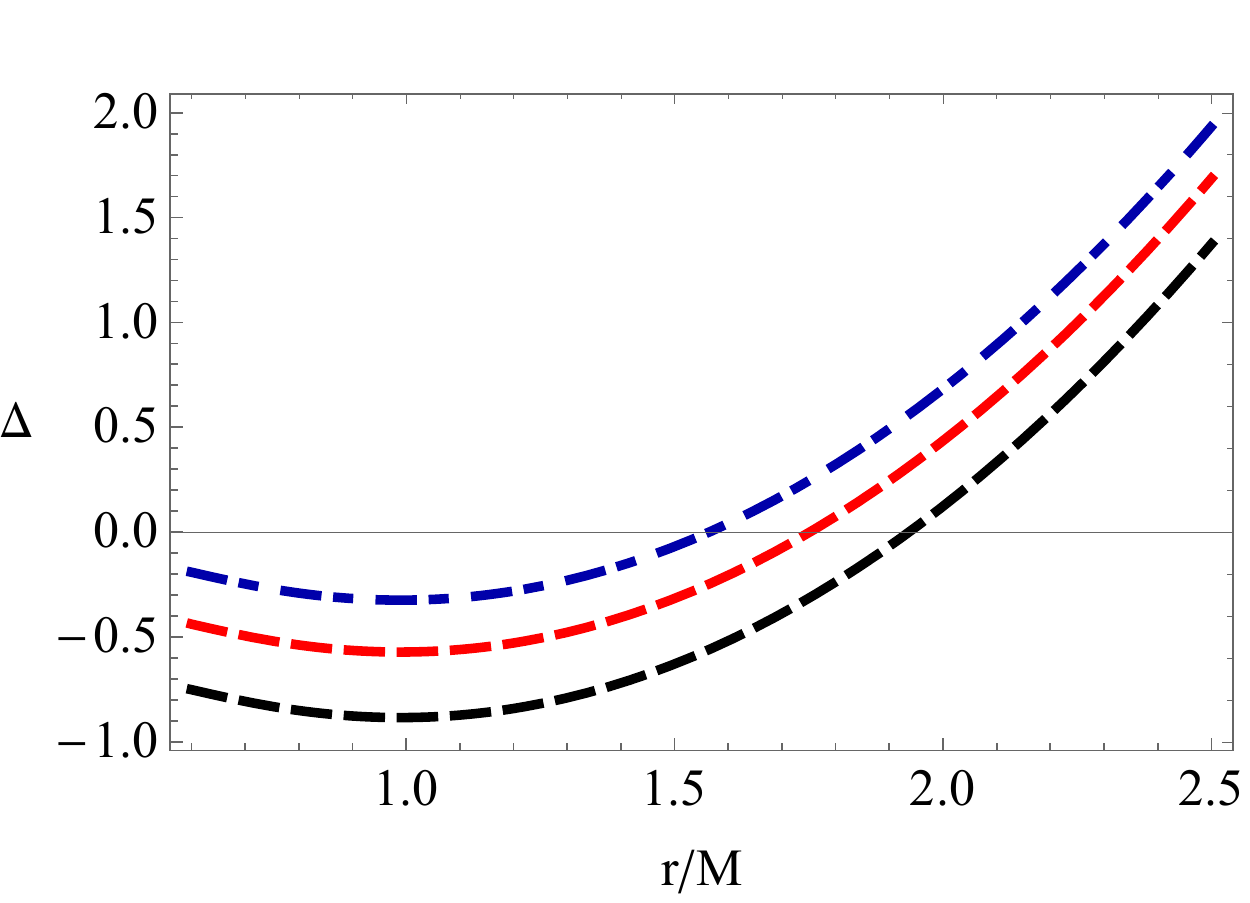}
		\includegraphics[scale=0.62]{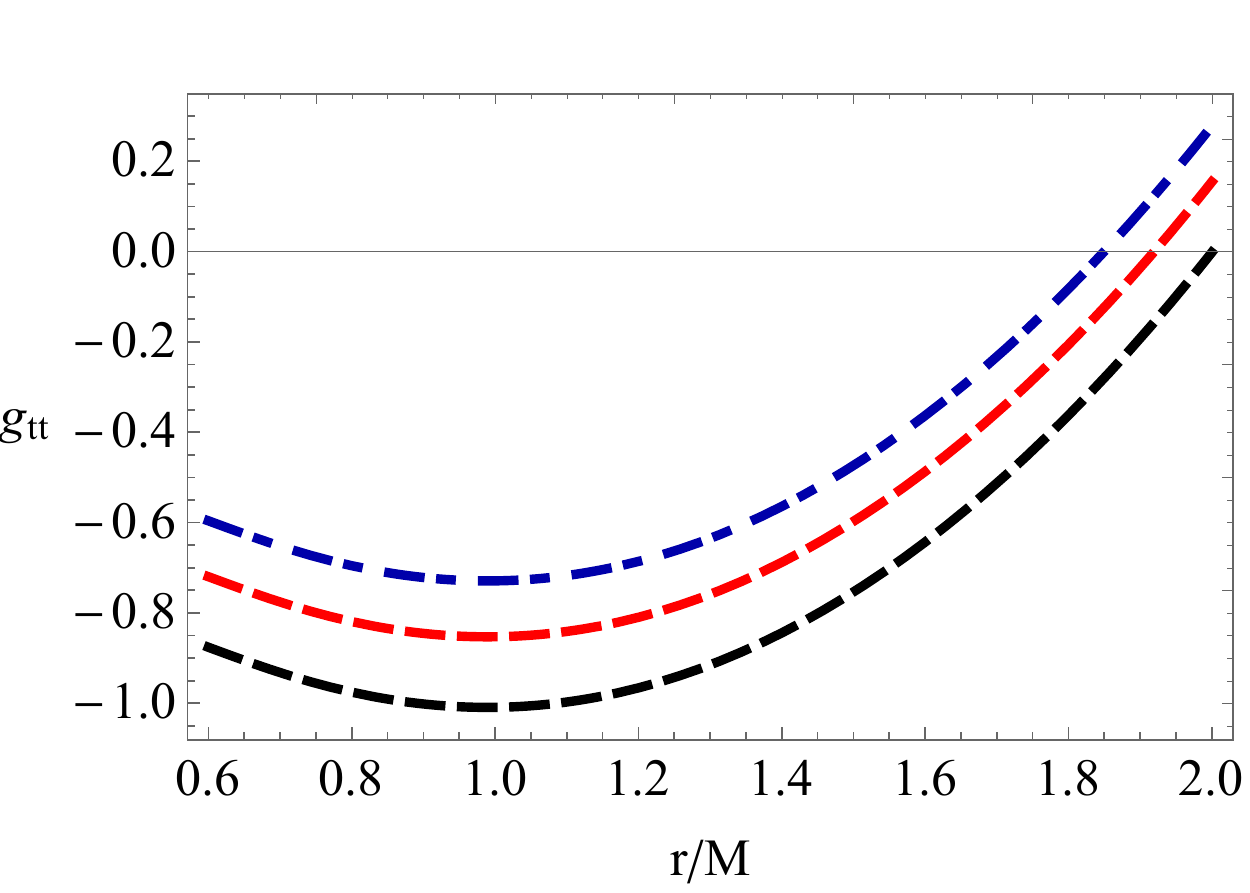}
		\caption{The plot of $\Delta$ and $g_{tt}$ as a function of raial coordinate for different values of angular momentum, black curve ($a=0.5$), red curve ($a=0.75$) and blue curve ($a=0.9$), respectively. We have set $b/M=0.5$ in both plots and $\theta=\pi/4$ in the second plot.  }\label{gttdelta}
	\end{figure*}

	\section{spinning Quantum Deformed Spacetime\label{secrqd}}
	The static and spherically symmetric quantum-deformed Schwarzschild spacetime metric as derived by KS is given by ~\cite{def}
	\begin{equation}
		ds^2 = -f(r) dt^2 + \frac{dr^2}{f(r)} + r^2\big(d \theta^2 + \sin^2\theta \,d\phi^2\big),
		\label{E:g_n}
	\end{equation}
	where
	\begin{equation}
		f(r) =   
		\left(1-\frac{b^{2}}{r^{2}}\right)^{1\over2}-\frac{2M}{r}.
		\label{E:f_n}
	\end{equation}
	Here \( b\in[0,\infty) \) and \( r\in[b,\infty) \). The case \( b = 0 \) reduces to the Schwarzschild BH in Schwarzschild coordinates. In Ref.~\cite{def}, $b^2$ was assumed positive for the purpose to shift the Schwarzschild singularity at $r=0$ to $r=b$ [the scalar curvature of~\eqref{E:g_n}-\eqref{E:f_n} diverges at $r=b$]. This, however, did not change the spacelike nature of the singularity. In this work we extend the domain of $b^2$ to include negative values. As we shall see, this will allow us to obtain perfect curve fitting of the particle QPO upper and lower frequencies to the observed frequencies for the best known microquasars.
	
	The counterpart spinning solution of the static metric~\eqref{E:g_n} is obtained via the Newman--Janis--Azreg-A\"{\i}nou algorithm (NJAA). Following~\cite{Azreg-Ainou:2014pra,Xu:2021lff} we arrive at
	the effective spinning BH  metric in Kerr-like coordinates
	\begin{eqnarray}\notag
		ds^2 &=& -\frac{\Delta}{\rho^2}(dt-a\sin^2\theta d\phi)^2+\frac{\rho^2}{\Delta}dr^2+\rho^2d\theta^2 \\
		\label{metric}	&+&\frac{\sin^2\theta}{\rho^2}[a dt-(r^2+a^2)d\phi]^2 ,  
	\end{eqnarray}
	where
	\begin{eqnarray}
		\Delta(r)&=& r^2f(r)+a^2,\\
		\rho^2(r,\theta)&=&r^2+a^2\cos^2\theta,
	\end{eqnarray}
	with $a$ is the specific angular momentum of the BH. After obtaining  the spinning BH solution \eqref{metric}, one can analyse the shape of the ergoregion and the BH horizons. For example, we can find the horizons by solving $\Delta=0$, or
	\begin{equation}
		r^2f(r)+a^2=0.
	\end{equation}
	The location for the ergo-surfaces can be obtained by solving $g_{tt}=0$, or
	\begin{equation}
		r^2f(r)+a^2 \cos^2\theta=0.
	\end{equation}

In Fig.~\ref{gttdelta} we plot $\Delta=0$ (left panel) and $g_{tt}=0$ (right panel) for different values of angular momentum but fixed value of $b$. Observe that the position of the event horizon decreases with the increase of the angular momentum. 

	\section{Relating QNMs and the shadow radius\label{secqnm}}
	
	In this section, we turn our attention to study the evolution of photons around the spinning quantum deformed BH. Let us start by writing the Hamilton-Jacobi equation
	\bqn
	\frac{\partial S}{\partial \lambda}=-\frac{1}{2}g^{\mu\nu}\frac{\partial S}{\partial x^\mu}\frac{\partial S}{\partial x^\nu},
	\eqn
	where $\lambda$ is an affine parameter. Next, we can write the Jacobi action $S$ in the standard form
	\begin{equation}
		S=\frac{1}{2}m_0^2\lambda-Et+L_z\phi+S_r(r)+S_\theta(\theta),
	\end{equation}
	 and use the fact that for photons one has $m_0=0$. Moreover, $E$ and $L_z$ are the energy and the angular momentum of the photon, respectively. The two functions $S_r(r)$ and $S_\theta(\theta)$ depend only on $r$ and $\theta$, respectively.
	
	Now by substituting the Jacobi action into the Hamilton-Jacobi equation, we obtain
	\bqn
	S_r(r)&=&\int^r\frac{\sqrt{R(r)}}{\Delta}dr,\\
	S_\theta(\theta)&=&\int^\theta\sqrt{\Theta(\theta)}d\theta,
	\eqn
	where
	\begin{eqnarray}
		R(r) &=&[(r^2+a^2)E-aL_z]^2-\Delta[ \mathcal{K}+(L_z-aE)^2]\\
		\Theta(\theta)&=& \mathcal{K}+(a^2E^2-L^2_z\csc^2\theta)\cos^2\theta,
	\end{eqnarray}
	with $ \mathcal{K}$ being the Carter constant. Then variation of the Jacobi action yields the following four equations of motion for the motion of photons
	\bqn \notag
	\rho^2\frac{dt}{d\lambda} &=&a(L_z-aE\sin^2\theta)+\frac{r^2+a^2}{\Delta}[(r^2+a^2)E -aL_z], \lb{YYY} \\\notag
	\rho^2\frac{d\phi}{d\lambda} &=&\frac{L_z}{\sin^2\theta}-aE+\frac{a}{\Delta}[(R^2+a^2)E-aL_z],\\\notag
	\rho^2\frac{dr}{d\lambda} &=&\sqrt{R(r)},\\
	\rho^2\frac{d\theta}{d\lambda}&=&\sqrt{\Theta(\theta)}.\lb{XXX}
	\eqn
	To this end, we can define the two impact parameters
	\bqn
	\xi=\frac{L_z}{E},\qquad \eta=\frac{\mathcal{K}}{E^2}.
	\eqn
	We can now determine the geometric shape of the shadow of the BH, towards this purpose we need to to use the unstable condition for circular geodesics
	\bqn
	R(r)=0,\qquad \frac{dR(r)}{dr}=0.\lb{ZZZ}
	\eqn
	%The geometric shape of the shadow is determined by the allowed values of $\xi$ and $\eta$ that fulfill these conditions. In general, the shape of the shadow depends on the rotation parameter $a$.
	
	\begin{figure*}[ht!]
		\centering
		\includegraphics[scale=0.6]{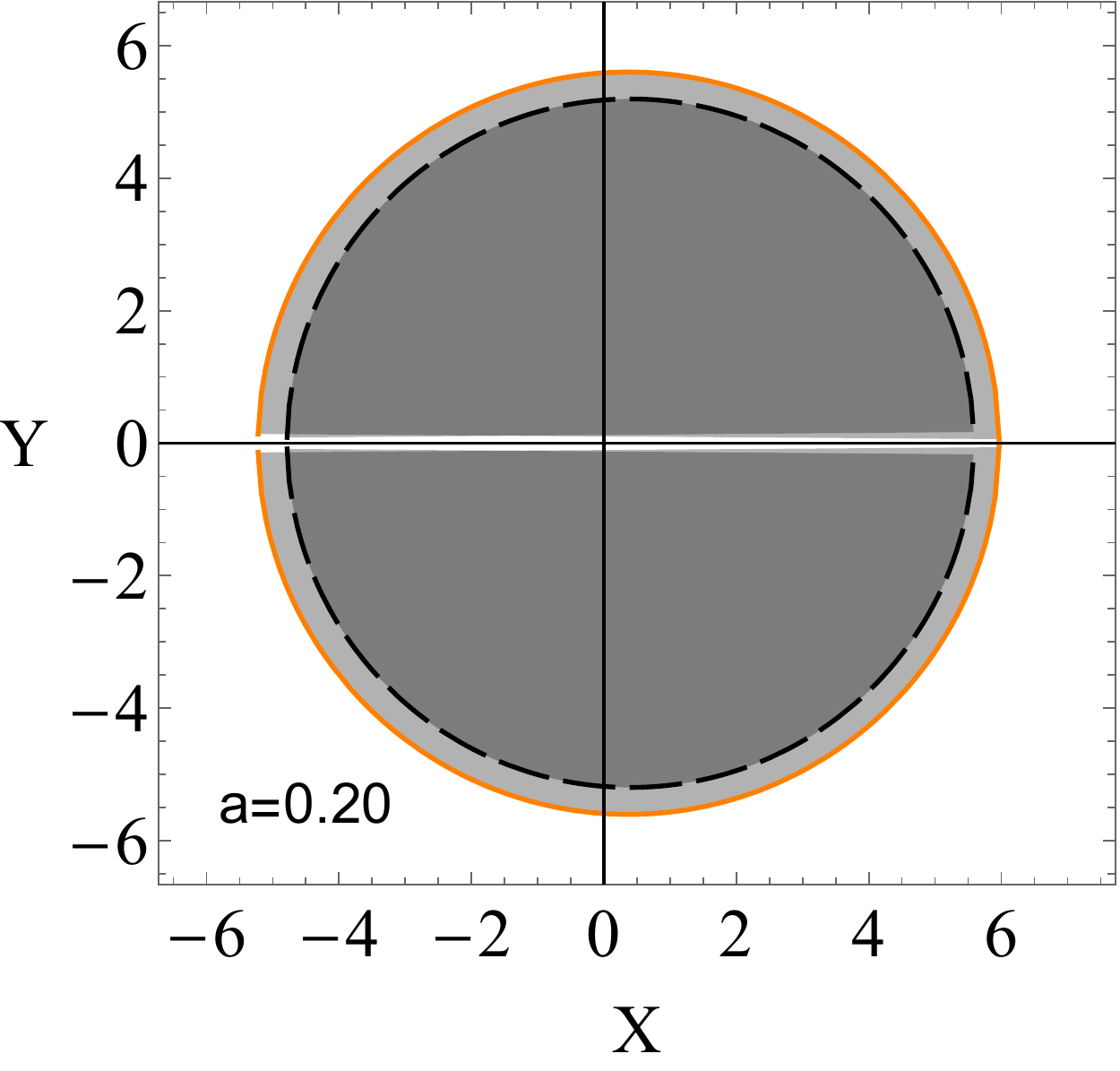}
		\includegraphics[scale=0.6]{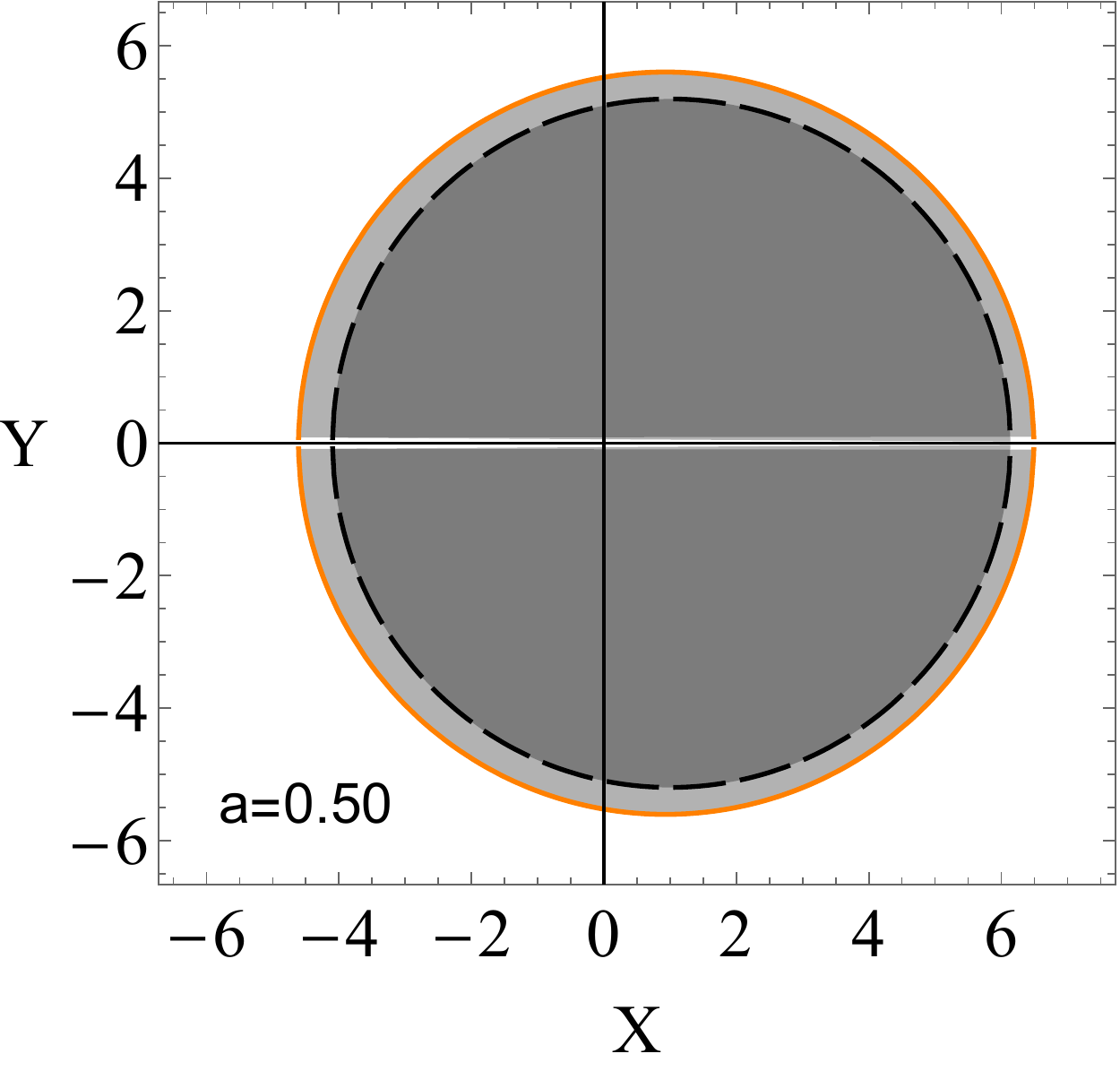}
		\includegraphics[scale=0.6]{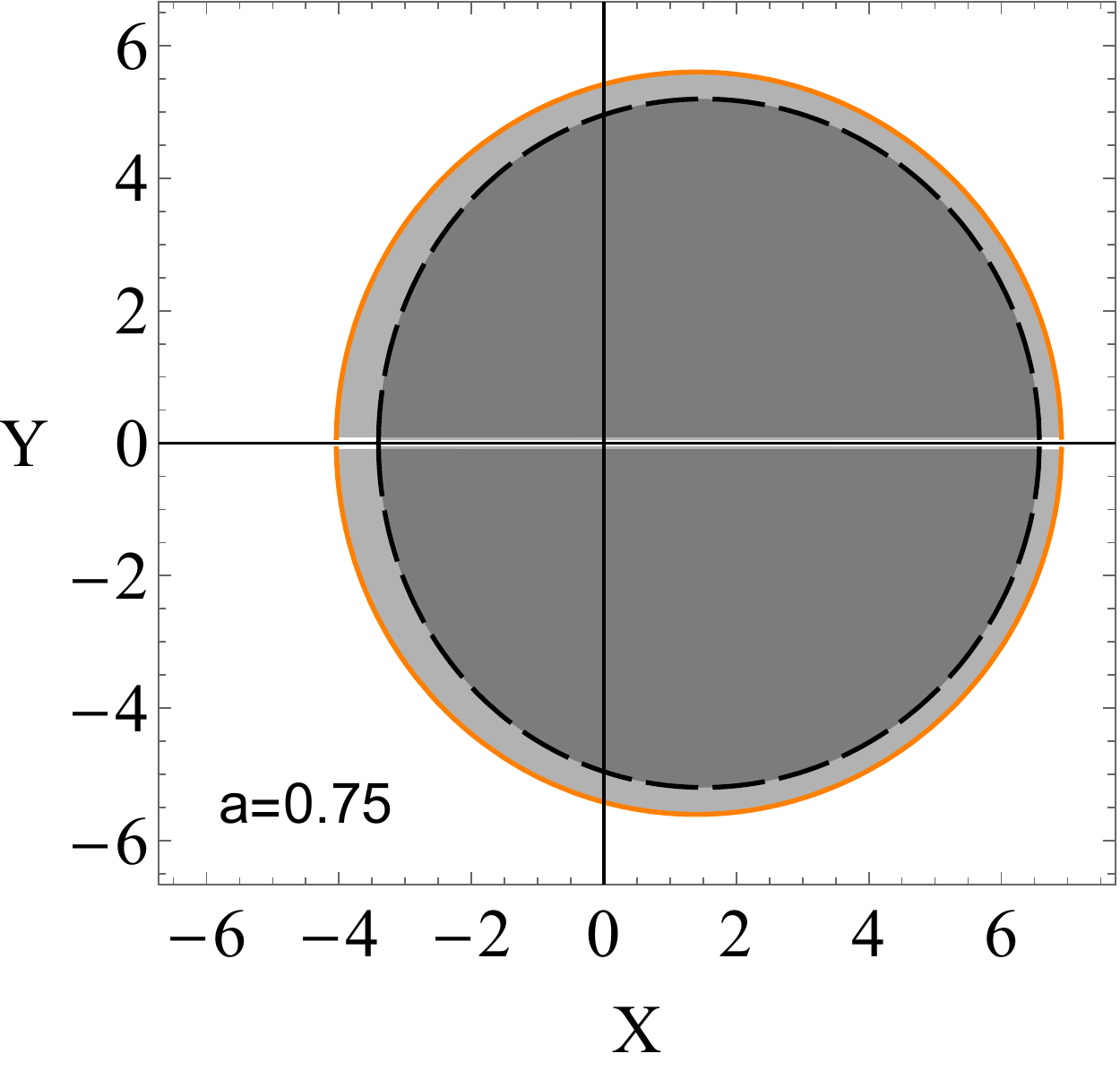}
		\includegraphics[scale=0.6]{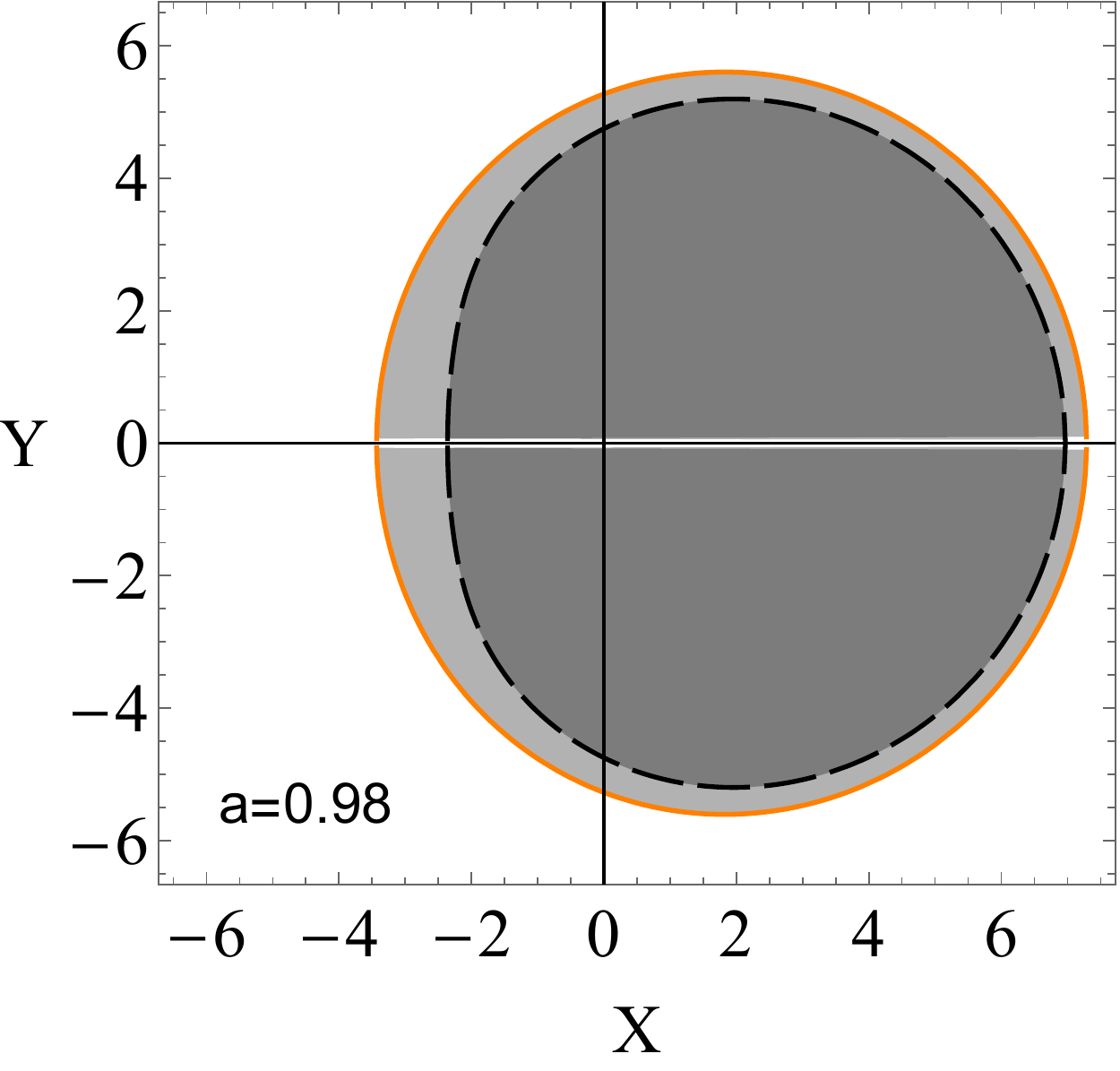}
		\caption{Shadows of spinning quantum deformed BHs using $b=1$ (measured in units $M$) and $\theta_o=\pi/2$. Note that the dashed curve
		represents the Kerr BH shadow.  }
	\end{figure*}
%	In this section, we {aim} to construct the shape of the shadow of the spinning quantum deformed BH.
In general, the photons emitted by a light source will be deflected when it passes by the BH because of the gravitational lensing effects. Some of the photons can reach the distant observer after being deflected by the BH, and some of them directly fall into the BH. The photons that cannot escape from the BH form the shadow of the BH in the observer's sky. The border of the shadow defines the apparent shape of the BH. To study the shadow, we adopt the celestial coordinates defined as follows:
	\begin{eqnarray}
		X&=&-\xi \csc{\theta_0},\\
		Y&=&\pm\sqrt{\eta+a^2 \cos^2\theta_0-\xi^2 \cot^2\theta_0}.
	\end{eqnarray}
	where  \cite{Shaikh:2019fpu}
	\begin{eqnarray}
		\xi(r)&=&\frac{\mathcal{X}_\text{0}\Delta'_\text{0}-2\Delta_\text{0}\mathcal{X}'_\text{0}}{a\Delta'_\text{0}},\\
		\label{eq:xi}
		\eta(r)&=&\frac{4a^2\mathcal{X}'^2_\text{0}\Delta_\text{0}-\left[\left(\mathcal{X}_\text{0}-a^2\right)\Delta'_\text{0}-2\mathcal{X}'_\text{0}\Delta_\text{0} \right]^2}{a^2\Delta'^2_\text{0}}.
		\label{eq:eta}
	\end{eqnarray}
	with $\mathcal{X}=r^2+a^2$. Here we note that the subscript ``$\text{0}$" indicates the above equations should be evaluated at $r=r_\text{0}$, which is a solution to~\eqref{ZZZ}. In Fig. 2 we show the shadow images of the Kerr deformed BH for different values of angular momentum. It is found that by increasing the parameter $b$ the shadow images increase compared to the Kerr vacuum solution. In this section we investigate the relation between the shadow and QNMs. It is well known that QNMs are modes related to ringdown phase of the BH and can be given in terms of the real and imaginary part as follows  $\omega=\omega_{\Re}- i \omega_{\Im}$. Recently it was argued that for static metrics there is a relation between the shadow radius and the real part of QNMs \cite{J1,c1} 
	\begin{equation}
		R_{s}=\frac{l+\frac{1}{2}}{\omega_{\Re}}.
	\end{equation}
	It it interesting that in Ref. \cite{Yang:2012he} it was shown that the QNM frequency of the Kerr BH in the eikonal limit reads
	\begin{eqnarray}
		\omega_{QNM}=(l+\frac{1}{2})\Omega_R-i \gamma_L \left(n+\frac{1}{2}\right)
	\end{eqnarray}
	with 
	\begin{eqnarray}
		\Omega_R=\Omega_{\theta}+\frac{m}{l+\frac{1}{2}}\Omega_{prec},
	\end{eqnarray}
	where $\Omega_{\theta}$ is the orbital frequency in the polar direction,
	$\Omega_{prec}$ is the Lense-Thiring precession frequency of the orbit plane, $\gamma_L$ is the Lyapunov exponent of the orbit, and $n$ is the overtone number. In what follows we are going to relate QNMs and the shadow radius measured by an observer located at some large distance from the BH. It is known that due to the rotation the BH shadow gets distorted as a result the shape of the shadow depends on the viewing angle. 
	
	\begin{itemize}
		\item Viewing angle: $\theta_0=\pi/2 $
	\end{itemize}
	In general there is no expression for the shadow, in what follows we shall consider the case $\theta_0=\pi/2 $. For simplicity, we shall also consider equatorial orbit which can be used to compute the typical shadow radius.  We are also going to use the following fact that the Lense-Thiring precession frequency is related to the orbital frequency and Keplerian frequency, namely, the Lense–Thirring precession frequency for prograde orbits in the limit of a small perturbation with respect to the equatorial plane is as follows
	\begin{eqnarray}
		\Omega_{prec}=\pm \Omega_{\phi}\mp \Omega_{\theta}\label{lt}
	\end{eqnarray}
	where 
	\begin{eqnarray}\label{Omegaf}
		\Omega_{\phi}=\frac{-\partial _r g_{t \phi }\pm \sqrt{\left(\partial _r g_{t \phi }\right)^2-(\partial _r g_{t t})( \partial _r g_{\phi  \phi })}}{\partial _r g_{\phi
				\phi }}.
	\end{eqnarray}
	
	In the case of spinning metric in Ref. \cite{J1} it was studied the case $m \pm l$, while in Ref. \cite{Y} it was argued that 
	\begin{eqnarray}
		\mathcal{K}+L_z^2 \simeq L^2-\frac{a^2 E^2}{2}\left(1-\frac{L_z^2}{L^2}\right).
	\end{eqnarray}
	We can rewrite the last equation as follows
	\begin{eqnarray}
		\eta+\xi^2 \simeq \frac{L^2}{E^2}-\frac{a^2 }{2}\left(1-\frac{L_z^2}{L^2}\right).\label{xi1}
	\end{eqnarray}
	
	Now if we use the following correspondence \cite{Yang:2012he} 
	\begin{eqnarray}
		L_z  & \Longleftrightarrow & m\\ 
		E & \Longleftrightarrow & \omega_{\Re}\\
		L & \Longleftrightarrow & l+\frac{1}{2}
	\end{eqnarray}
	where the real part of QNMs is also given by $\omega_{\Re}=L \Omega_R$. In the eikonal limit $m=l>>1$, we have, hence if we introduce $\mu=m/(l+1/2)=1$ as a result, by means of Eq. \eqref{lt} with  a positive sign before the orbital frequency 
	\begin{eqnarray}
		\Omega_{prec}=\Omega_{\phi}-\Omega_{\theta}
	\end{eqnarray}
	as a result we obtain
	\begin{eqnarray}
		\Omega_R=\Omega_{\theta}+\Omega_{prec}=\Omega_{\phi},
	\end{eqnarray}
	where the term $\Omega_{\theta}$ cancels out. It means that the real part of QNMs is related to the Kepler frequency given by 
	\begin{eqnarray}
		\omega^{\pm}_{\Re}=(l+\frac{1}{2})\frac{-\partial _r g_{t \phi }\pm \sqrt{\left(\partial _r g_{t \phi }\right)^2-\partial _r g_{t t} \partial _r g_{\phi  \phi }}}{\partial _r g_{\phi
				\phi }}\label{rpart}
	\end{eqnarray}
	which is valid in the limit $m=l>>1$. Alternatively, we can consider the mode $m=-l$, it follows that $\mu=-1$, where we can use again Eq. \eqref{lt} with negative sign before the Kelperian frequency and we arrive again at Eq. \eqref{rpart}. In this case, we can use the following definition to specify the shadow radius \cite{Feng,J1}
	\begin{eqnarray}
		R_s:=\frac{1}{2}\left(\xi^+(r_{0}^+)-\xi^-(r_{0}^-)\right),\label{de1}
	\end{eqnarray}
	provided $\eta(r_{0}^{\pm})=0$. From Eq. \eqref{xi1} one can obtain 
	\begin{eqnarray}
		\xi^{\pm}=\pm \sqrt{\frac{(l+\frac{1}{2})^2}{\omega^2_{\Re}(r_{0}^{\pm})}-\frac{a^2}{2}(1-\mu^2)}.
	\end{eqnarray}
	
	With these equations in hand, it follows that
	\begin{eqnarray}\notag
		R_s&=&\frac{1}{2}\sqrt{\frac{(l+\frac{1}{2})^2}{\omega^2_{\Re}(r_{0}^+)}-\frac{a^2}{2}(1-\mu^2)}\\
		&&+  \frac{1}{2}\sqrt{\frac{(l+\frac{1}{2})^2}{\omega^2_{\Re}(r_{0}^-)}-\frac{a^2}{2}(1-\mu^2)}.
	\end{eqnarray}
	The correspondence is  precise if we consider the eikonal limit, that is, if we set $\mu=\pm 1$ (namely $ [(m=\pm l)]$, yielding
	\begin{equation}
		R_s(\mu=\pm 1)=\frac{l+\frac{1}{2}}{2}\left(\frac{1}{\omega_{\Re}(r_{0}^+)}+\frac{1}{\omega_{\Re}(r_{0}^-)}\right).
	\end{equation}

	\begin{figure}[h]
		\centering
		\includegraphics[width=0.46\textwidth]{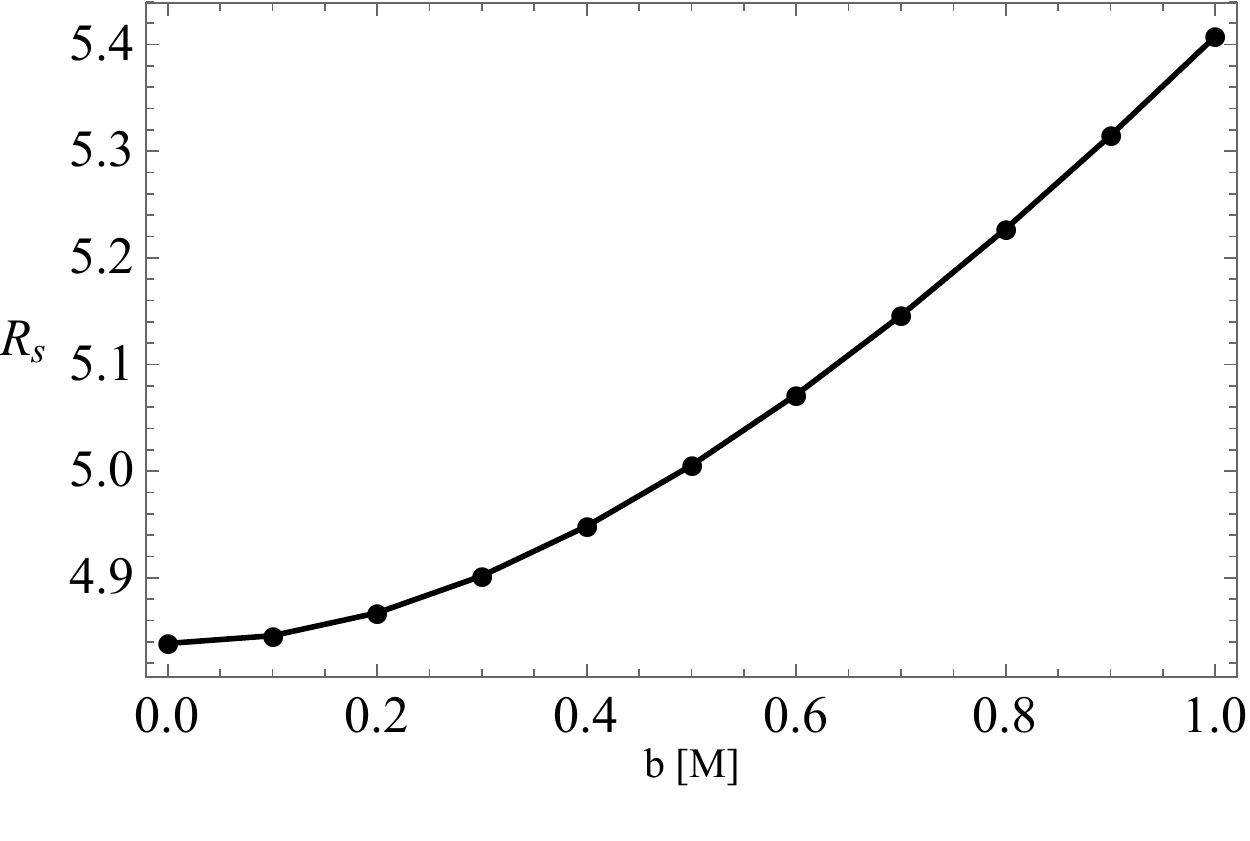}
		\caption{Typical shadow radius by varying the parameter $b$, using a viewing angle $\theta_0=\pi/2$. We have used $a=0.9$.}
	\end{figure}
	
	\begin{figure}[h]
		\centering
				\includegraphics[width=0.46\textwidth]{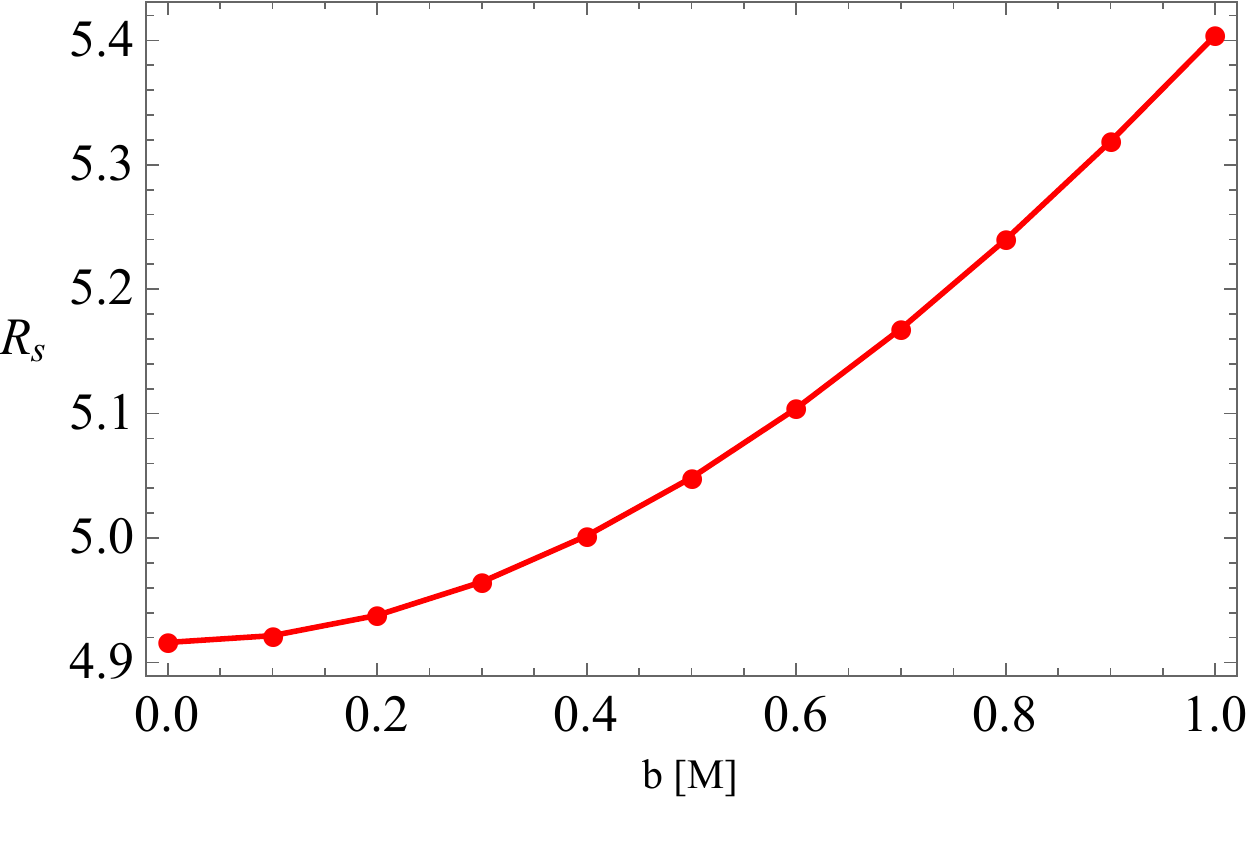}
		\caption{Typical shadow radius by varying the parameter $b$, using a viewing angle $\theta_0=0$ (or equivalently $\theta_0=\pi$). We have used $a=0.9$.}
	\end{figure}
	
		\begin{figure*}
   \centering
   \includegraphics[scale=0.65]{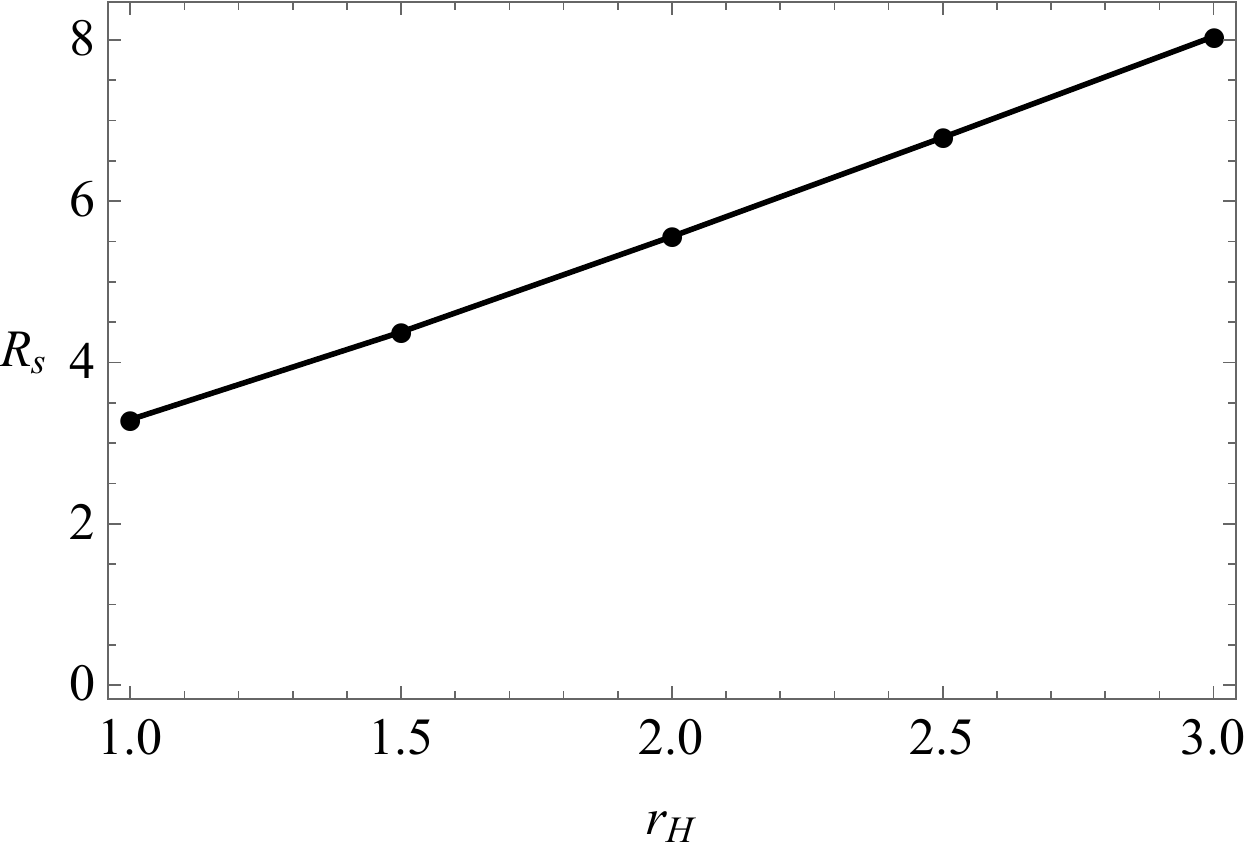}
     \includegraphics[scale=0.65]{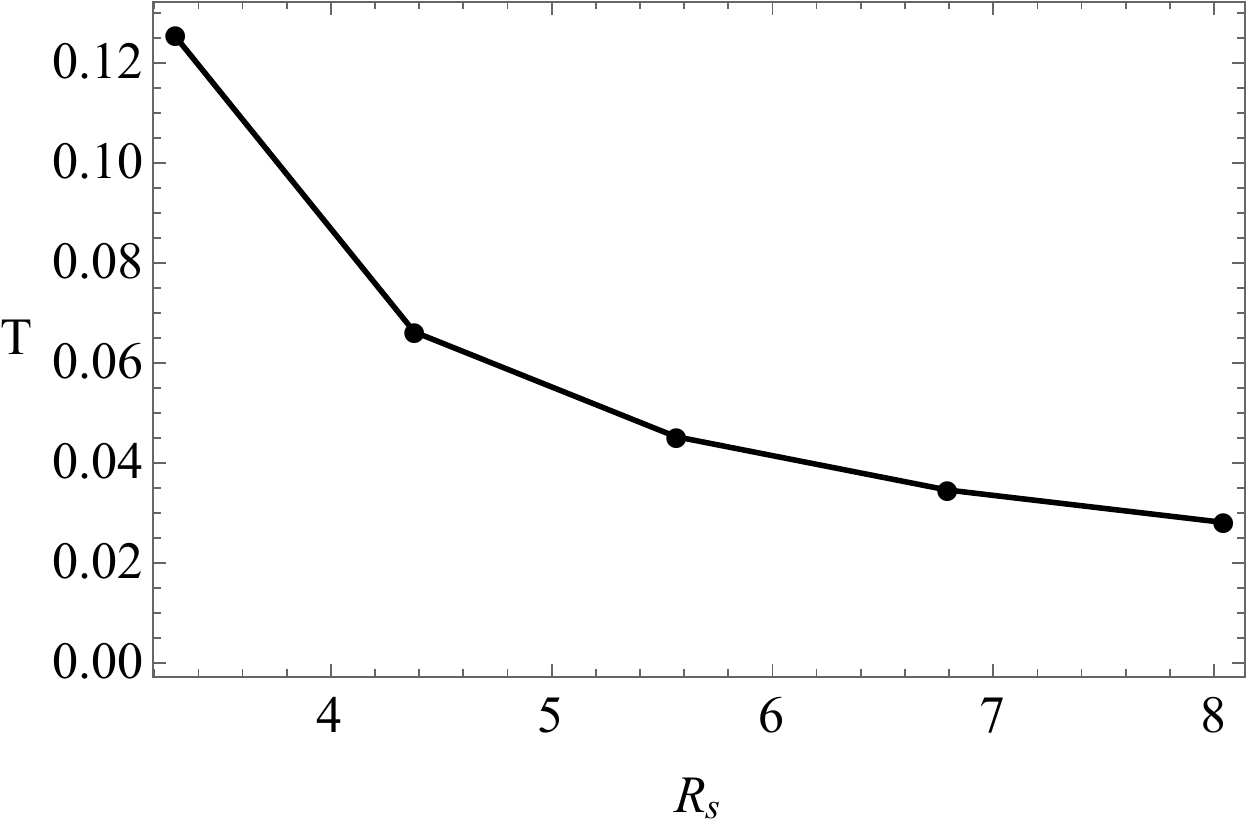}
    \caption{ Left panel: The typical shadow radius of the spinning quantum deformed BH viewed in $\theta_o=\pi/2$  as a function of the event horizon radius. Right panel: The Hawking temperature of the spinning quantum deformed BH as a function of typical shadow radius.  We have set $a/M=0.65$ and $b/M=0.5$.  }
\end{figure*}

\begin{figure*}
    \centering
    \includegraphics[scale=0.65]{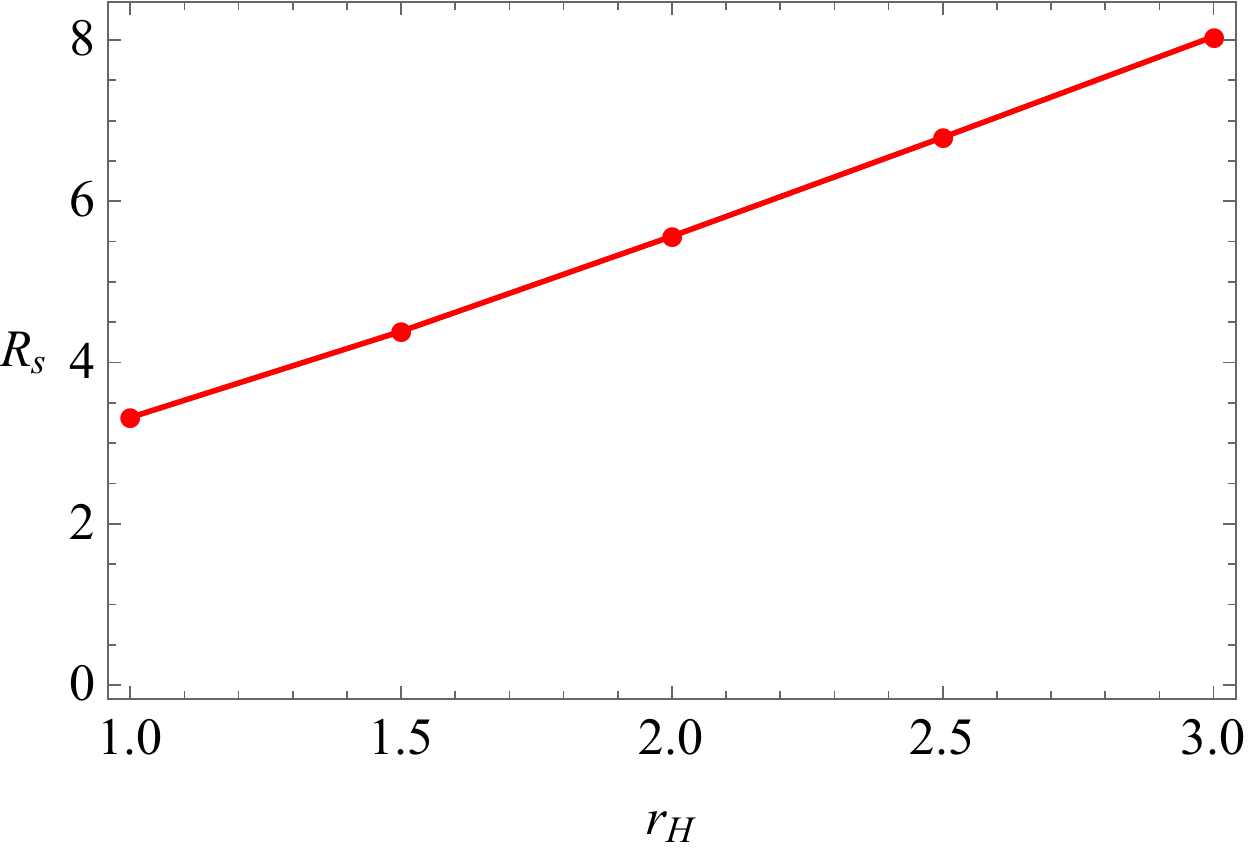}
     \includegraphics[scale=0.65]{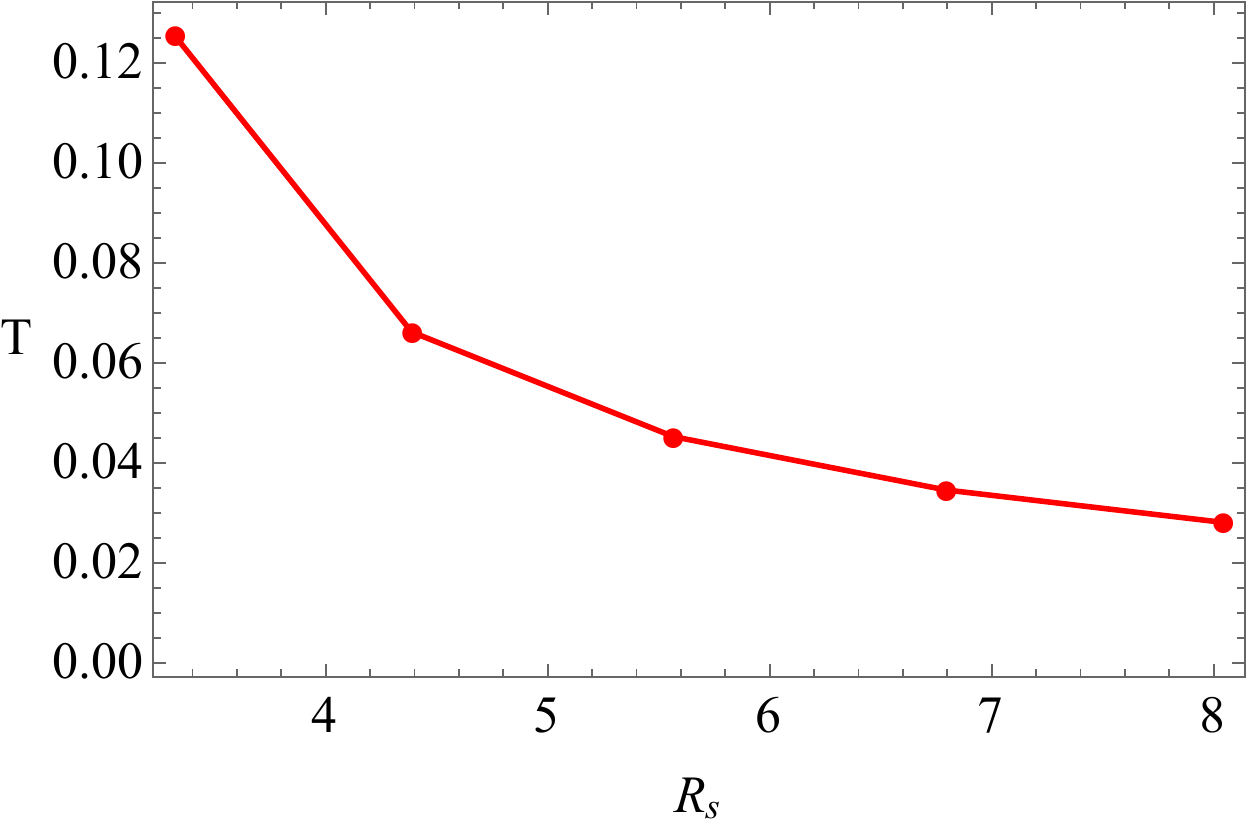}
    \caption{  Left panel: The typical shadow radius of the spinning quantum deformed BH viewed in $\theta_o=0$  as a function of the event horizon radius. Right panel: The Hawking temperature of the spinning quantum deformed BH as a function of typical shadow radius.  We have set $a/M=0.65$ and $b/M=0.5$ . }
\end{figure*}
	
\begin{figure*}
    \centering
    \includegraphics[scale=0.65]{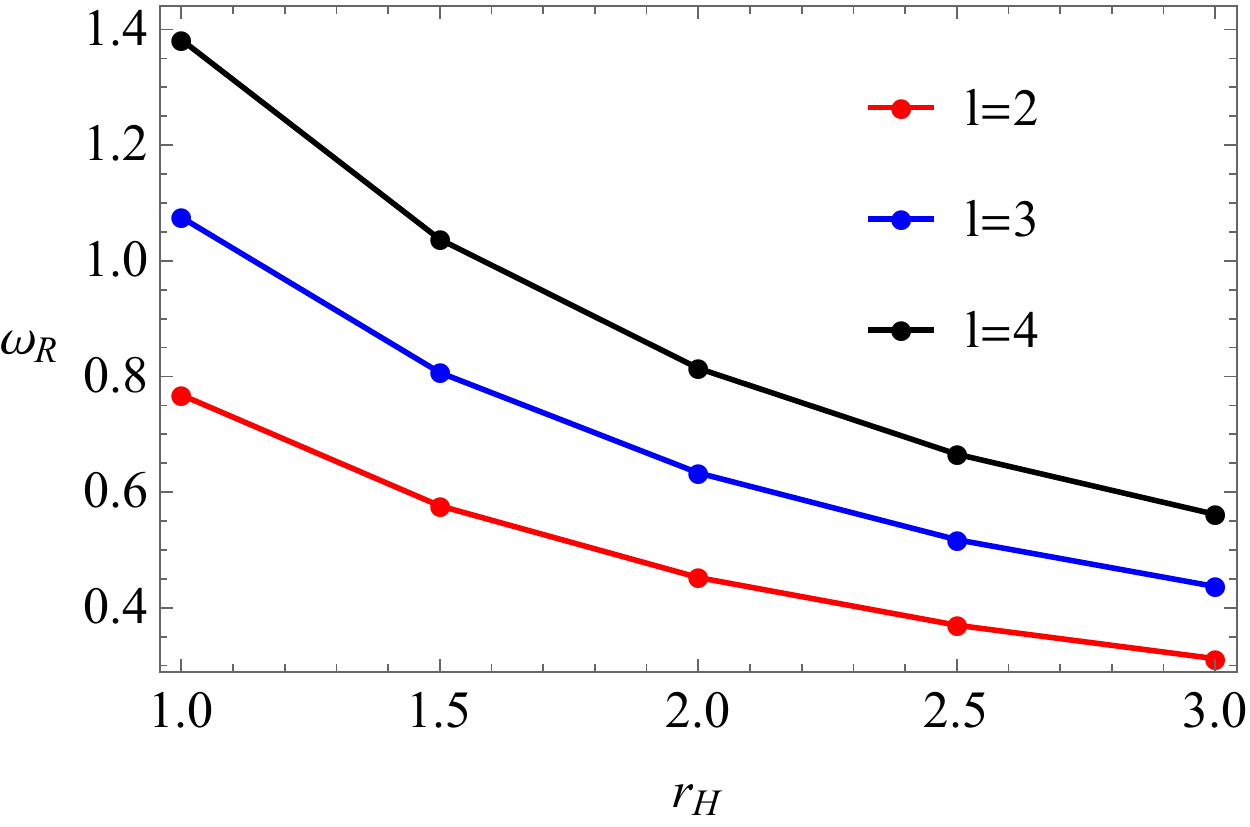}
     \includegraphics[scale=0.65]{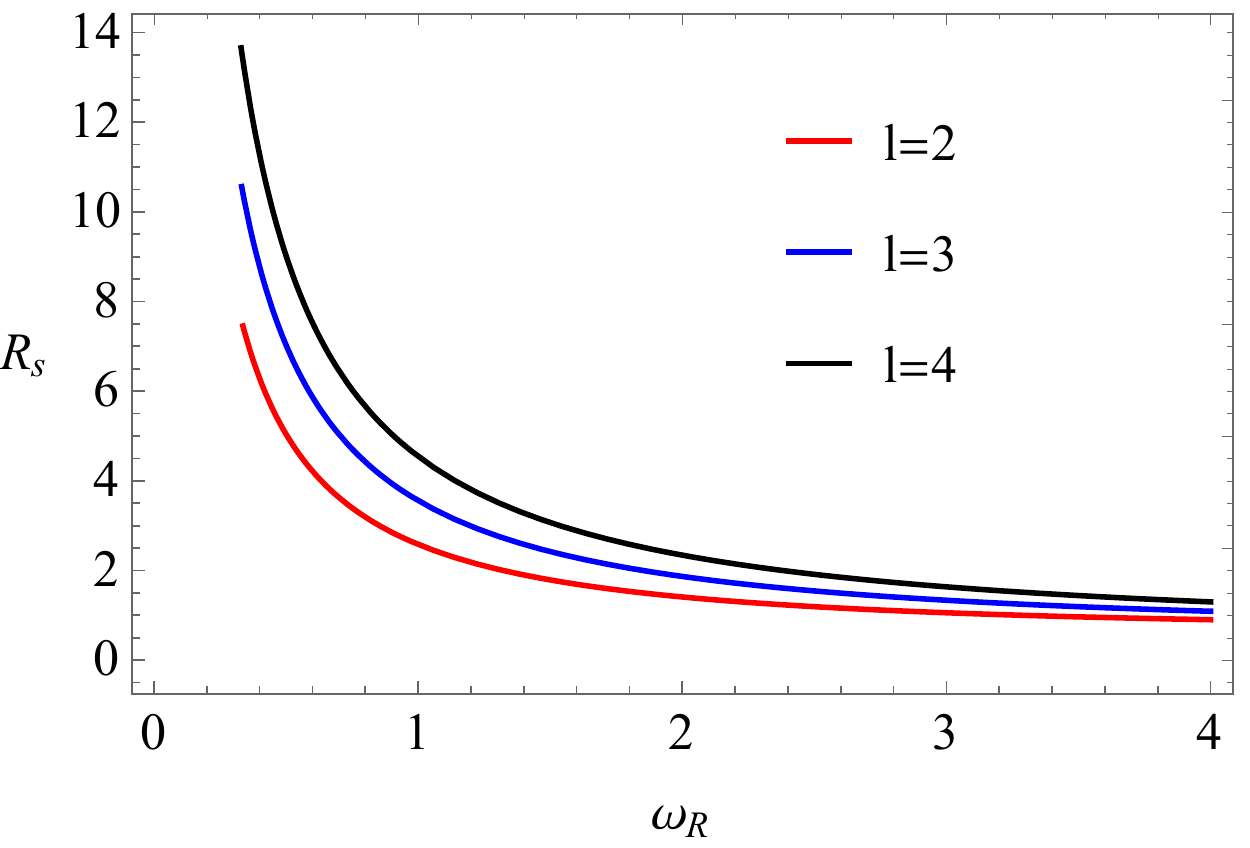}
    \caption{Left panel: The real part of QNMs as a function of the horizon radius for different $l$. We have set $a/M=0.65$ and $b/M=0.5$. Right panel: The typical shadow radius (Eq. (71)) as a function of the real part of QNMs with $a/M=0.65$. }
\end{figure*}

	The last equation  coincides with the expression obtained in \cite{J1}, under the definition $\omega^+_{\Re}=-\omega^-_{\Re}=\omega_{\Re}$. Now by means of Eq. \eqref{rpart} and the metric functions in the equatorial plane:
	\begin{eqnarray}
		g_{t\phi}&=&-a (1-f(r)),\\
		g_{tt}&=&-f(r),\\
		g_{\phi \phi}&=& \left((r^2+a^2)^2-a^2[r^2 f(r)+a^2]\right)/r^2,
	\end{eqnarray}
it is possible to arrive at the following result for the typical shadow radius
		\begin{equation}\notag
			R_s=\frac{1}{2}\Big(\frac{\partial _r g_{\phi\phi }|_{r_0^+}}{-\partial _r g_{t \phi }|_{r_0^+} + \sqrt{\left(\partial _r g_{t \phi }\right)^2|_{r_0^+}-\partial _r g_{t t}|_{r_0^+} \partial _r g_{\phi  \phi }|_{r_0^+}}}
			\end{equation}
			\begin{equation}
			+\frac{\partial _r g_{\phi\phi }|_{r_0^-}}{-\partial _r g_{t \phi }|_{r_0^-} - \sqrt{\left(\partial _r g_{t \phi }\right)^2|_{r_0^-}-\partial _r g_{t t}|_{r_0^-} \partial _r g_{\phi  \phi }|_{r_0^-}}}\Big).
		\end{equation}
	However, after some algebraic manipulation we can rewrite the above equation as follows
%	\begin{widetext}
		\begin{eqnarray}
			R_s&=&\frac{1}{2}\Big(\frac{\partial _r g_{t \phi}|_{r_0^-}}{\partial _r g_{t t}|_{r_0^-}}-\frac{\partial _r g_{t \phi}|_{r_0^+}}{\partial _r g_{t t}|_{r_0^+}}\Big)- \frac{1}{2}\left(\mathcal{A}+\mathcal{B}\right).
		\end{eqnarray}
%	\end{widetext}
	where 
	\begin{eqnarray}
	\mathcal{A}&=&\frac{\sqrt{(\partial _r g_{t \phi})^2|_{r_0^+}-\partial _r g_{\phi \phi}|_{r_0^+}\, \partial _r g_{t t}|_{r_0^+}}}{\partial _r g_{t t}|_{r_0^+}}\\
	\mathcal{B}&=&\frac{\sqrt{(\partial _r g_{t \phi})^2|_{r_0^-}-\partial _r g_{\phi \phi}|_{r_0^-}\, \partial _r g_{t t}|_{r_0^-}}}{\partial _r g_{t t}|_{r_0^-}}
	\end{eqnarray}
	From the last equation one can check that the first two terms cancel out since
	\begin{eqnarray}
			\frac{\partial _r g_{t \phi}|_{r_0^{\pm}}}{\partial _r g_{t t}|_{r_0^{\pm}}}=-a.
		\end{eqnarray}
	As a result we obtain a simple equation
	\begin{equation}
		R_s=\frac{\sqrt{2}}{2}\left(\sqrt{\frac{ r_0^{+}}{f'(r)|_{r_0^{+}}}}+\sqrt{\frac{ r_0^{-}}{f'(r)|_{r_0^{-}}}}\right).\label{rs}
	\end{equation}
	The last equation is nothing but the result which was obtained previously in Ref. \cite{J1,G}, where  the points $r_0^{\pm}$ were determined by solving
	\begin{equation}
		r_0^2-\frac{2 r_0}{f'(r)|_{r_0^{\pm}}}f(r_0)\mp 2 a \sqrt{\frac{2 r_0}{f'(r)|_{r_0^{\pm}}}}=0.
	\end{equation}
	
	We can derive an alternative expression for the typical shadow radius, using $\eta(r_0^{\pm})=0$, from Eq. \eqref{eq:eta} it follows that
	\begin{eqnarray}\notag
		\mathcal{X}(r)\Delta'(r)&-& 2\Delta(r)\mathcal{X}'(r)|_{r_0^{\pm}}= \pm 2 a \mathcal{X}'(r)\sqrt{\Delta(r)}|_{r_0^{\pm}}\\
		&+& a^2 \Delta'(r)|_{r_0^{\pm}}.
	\end{eqnarray}
	Further making use of the definition \eqref{de1}, it follows that 
	\begin{eqnarray}
		R_s=\frac{ \mathcal{X}'(r_0^+) }{\Delta'(r_0^+)}\sqrt{\Delta(r_0^+)}+\frac{ \mathcal{X}'(r_0^-) }{\Delta'(r_0^-)}\sqrt{\Delta(r_0^-)}.
	\end{eqnarray}
	The last equation is equivalent to Eq. \eqref{rs} and gives same results. The QNMs frequency for $\mu \neq 1$ in the quantum deformed Kerr case  can be approximated by using \footnote{See Eq.~\eqref{Omega} in section VI for details.}
	\begin{eqnarray}
		\Omega_\theta \simeq \frac{(4 r + b^2/M)\sqrt{M}}{4 r^{5/2}}+\frac{Ma}{r^3},
	\end{eqnarray}
	Hence the real part of QNMs reads 
	\begin{equation}
		\omega_{\Re}^{\pm} \simeq (l+\frac{1}{2})\left(\frac{(4 {r_0^{\pm}} + b^2/M)\sqrt{M}}{4 {r_0^{\pm}}^{5/2}}\pm \frac{|m|}{l+\frac{1}{2}} \frac{2 Ma}{{r_0^{\pm}}^3}  \right).
	\end{equation}
	
	As a special case if we set $a=b=0$, we obtain the QNMs frequency for the static BH given by \cite{J1}
	\begin{eqnarray}
		\omega_{\Re}= (l+\frac{1}{2}) \sqrt{\frac{M}{r_0^3}}=(l+\frac{1}{2})\frac{1}{R_s}
	\end{eqnarray}
	where $r_0=3M$ and $R_s=3 \sqrt{3} M$ are the photon radius and the shadow radius. As we elaborated above, in the eikonal limit we can use Eq. \eqref{rpart} which can be further simplified as follows 
	\begin{eqnarray}
		\omega_{\Re}^{\pm}=(l+\frac{1}{2}) \frac{1}{a \pm \sqrt{\frac{ 2 r_0^{\pm}}{f'(r)|_{r_0^{\pm}}}} }.
	\end{eqnarray}
	
	The last equation reduces to the Kerr BH case as reported earlier in \cite{mash}, with the only difference that here we shall use the scaling $l \to l+1/2$, which gives better results.   In Fig. 3 we show the typical shadow radius as a function of the parameter $b$, where it is shown that the shadow increases with the increase of the value of $b$, for a fixed angular momentum. 
	\begin{table*}[tbp]
		\begin{tabular}{|l|l|l|l|}
			\hline
			\multicolumn{1}{|c|}{ } &  \multicolumn{2}{c|}{ Equatorial modes  } &   \multicolumn{1}{c|}{  Polar modes  }
			\\ %& \multicolumn{1}{c|}{} \\
			\hline
			$l$ & \quad $\omega_{\Re}^+$  & \quad $\omega_{\Re}^-$ & \quad $\omega_{\Re}$\\
			%& $R_s$  \\
			\hline
			1 & 0.2074765304 & 0.3542365088  &  0.2879754264 \\
			2 & 0.3457942174 & 0.5903941813 &   0.4799590440 \\
			3 & 0.4841119044 & 0.8265518538 &  0.6719426616 \\
			4 & 0.6224295913 & 1.0627095260 &  0.8639262792  \\
			5 & 0.7607472783 & 1.2988671990 &  1.0559098970 \\
			6 & 0.8990649653 & 1.5350248710 &  1.2478935140  \\
			7 & 1.0373826520 & 1.7711825440 &   1.4398771320  \\
			8 & 1.1757003390 & 2.0073402160 &  1.6318607500 \\
			9 & 1.3140180260 & 2.2434978890 &  1.8238443670 \\
			10 & 1.4523357130 & 2.4796555610  &  2.0158279850  \\
			\hline
		\end{tabular}
		\caption{ \label{table4} Numerical values of the real part of QNMs for equatorial modes and polar modes. We have set $a=0.5 [M]$ and $b=0.5 [M]$.}
	\end{table*}
	
	\begin{itemize}
		\item Viewing angle: $\theta_0=0$ \&  $\theta_0=\pi$
	\end{itemize}
	We are now going to consider the polar orbit $\theta=0$ and, in such a case, it is natural to compute the viewing angle for the observer: $\theta_0=0$ \&  $\theta_0=\pi$. From the celestial coordinates we can obtain
	\begin{eqnarray}
		X^2+Y^2=\eta+\xi^2+a^2 \cos^2\theta_0.
	\end{eqnarray}
	For a viewing angle $\theta_0=0$, the shadow remains a round disk hence we can use the following definition \cite{Feng}
	\begin{eqnarray}
		R_s:=\sqrt{a^2+\eta(r_0)},\,\,\,\text{with}\,\,\xi(r_0)=0.
	\end{eqnarray}
	From the condition $\xi(r_0)=0$, we get
	\begin{equation}
		\mathcal{X}(r_0)\Delta'(r_0)-2\Delta(r_0)\mathcal{X}'(r_0)=0,
	\end{equation}
	thus, we obtain 
	\begin{equation}
		R_s=\frac{2 \mathcal{X}'(r_0) }{\Delta'(r_0)}\sqrt{\Delta(r_0)}.\label{eq55}
	\end{equation}
		%\begin{equation}
	%	R_s=\frac{4 r_0 }{2 r_0 f(r_0)+r_0^2 f'(r_0)}\sqrt{r_0^2 f(r_0)+a^2}.
	%\end{equation}
	To our best knowledge the last equation has not been reported before in the literature. We can rewrite the last equation in terms of the function $f(r)$ as follows 
	\begin{equation}
		R_s=\frac{4 r_0 }{2 r_0 f(r_0)+r_0^2 f'(r_0)}\sqrt{r_0^2 f(r_0)+a^2}.
	\end{equation}
	In fact, if we consider the static limit $(a=0)$, then Eq. (56) gives 
	\begin{equation}
		r_0 f'(r_0)-2 f(r_0)=0.
	\end{equation}
	Hence, the shadow radius gives the well-known result given by
	\begin{equation}
		R_s|_{a=0}=\frac{r_0}{\sqrt{f(r_0)}}.
	\end{equation}
	
	For the polar orbit, we have zero azimuthal angular momentum, $L_z=0$, along with the conditions for the existence of circular geodesics i.e., $\dot{r}^2$, from Eq. (12) we obtain
	\begin{equation}
		(r^2+a^2)^2-[r^2f(r)+a^2]R_s^2=0,\label{def2}
	\end{equation}
	and
	\begin{equation}
		4 r (r^2+a^2)-2 r f(r)R_s^2-r^2f'(r)R_s^2=0,
	\end{equation}
	where $R_s^2=\mathcal{K}/E^2-a^2$ (see, \cite{Dolan}) has been used. From Eq. \eqref{def2} we obtain 
	\begin{equation}
		R_s^{\pm}=\pm \frac{a^2+r^2}{\sqrt{r^2 f(r)+a^2}}|_{r_0},
	\end{equation}
	In this case, the typical shadow radius can be defined as $R_s\equiv(R_s^+-R_s^-)/2$, hence we obtain
	\begin{equation}
		R_s=\frac{a^2+r^2}{\sqrt{r^2 f(r)+a^2}}|_{r_0},\label{eq61}
	\end{equation}
	where $r_0$ can be determined by solving the algebraic equation
	\begin{eqnarray}
		(a^2+r_0^2)^2-\frac{4 [r_0^2 f(r_0)+a^2](a^2+r_0^2)}{r_0 f'(r_0)+2 f(r_0)}=0.
	\end{eqnarray}
	Note that the expression for the typical shadow radius  given by \eqref{eq61} should lead to same result as the Eq. \eqref{eq55}.
	On the other hand, based on the relation between the shadow radius with the real part of QNMs it follows
	\begin{eqnarray}
		R_s=\sqrt{\frac{(l+1/2)^2}{\omega^2_{\Re}(r_0)}+\frac{a^2}{2}\left(1+\mu^2\right)}.
	\end{eqnarray}
	That is, in the eikonal limit with $\mu=\pm 1$, the real part of QNMs can be expressed as follows
	\begin{equation}
		\omega_{\Re}=(l+\frac{1}{2})\sqrt{\frac{r_0^2 f(r_0)+a^2}{r_0^4+r_0^2 a^2 (2-f(r_0))}}.
	\end{equation}
	
		Fig. 4 represents plot of the typical shadow radius for for the viewing angle $\theta_0=0$ (and equivalently $\theta_0=\pi$)  as a function of the parameter $b$. Observe that the shadow radius decreases when $b$ increases. From Figs. 3-4 we can see that the expression for the typical shadow radius behaves almost the same for both viewing angles, although for the viewing angle $\theta_0=0$ the value of the $R_s$ is slightly greater compared to the case $\theta_0=\pi/2$.  Finally in Table I, we have presented the numerical values for the QNMs for the case of equatorial modes and polar modes. There we can see that with the increase of $l$ the QNMs frequency increases.
	
\section{Thermodynamical stability of the spinning quantum deformed BH and BH shadow}
In this section we shall use the expression for the typical shadow radius to investigate the phase transition of the spinning quantum deformed adopting the method in Ref. \cite{Zhang:2019glo}.  
In particular, depending on the sign of the specific heat $C=T(\partial S/\partial T)$, we can have the case of stable ($C>0$) and unstable ($C<0$) BHs from a thermodynamical point of view. Note here that $T$ denotes the Hawking temperature associated with the event horizon of the BH. Using the relation for the entropy $dS/dr_H>0$, and the relation for the Hawking temperature  \cite{Zhang:2019glo} 
\begin{equation}
\frac{\partial T}{\partial r_{H}}=\frac{\partial T}{\partial R_{s}}\frac{\partial R_{s}}{\partial r_{H}},
\end{equation}
yields three cases
\begin{equation}\label{sphereone}
\frac{\partial T}{\partial r_{H}}>0,~~\frac{\partial T}{\partial r_{H}}=0,~~\frac{\partial T}{\partial r_{H}}<0,
\end{equation}
where $r_H$ is the event horizon of the BH. Moreover we have three additional cases
\begin{equation}\label{spheretwo}
\frac{\partial T}{\partial R_{s}}>0,~~\frac{\partial T}{\partial R_{s}}=0,~~\frac{\partial T}{\partial R_{s}}<0.
\end{equation}

In Fig. 5 and Fig. 6,  we have shown the change of the typical shadow radius for the viewing angle $\theta_0=\pi/2$ and $\theta_0=0$, in terms of the event horizon radius and found that the condition $\partial R_s/\partial r_H>0$ is indeed satisfied for both cases. Moreover, here we have also plotted the Hawking temperature of the BH horizon against the shadow radius. From these plots we confirmed that $\partial T/\partial R_s<0$, i.e. the spinning quantum deformed BH is thermodynamically unstable. Using the relation between the shadow radius and the real part of QNMs, we can formulate the problem of the stability in terms of the QNMs. Take for example the expression for the shadow radius (66) in the eikonal  limit, which reads
\begin{eqnarray}\label{R2}
		R_s^2=\frac{(l+1/2)^2}{\omega^2_{\Re}}+a^2.
	\end{eqnarray}
Differentiating the last equation with respect to $\omega_{\Re}$, we obtain
\begin{eqnarray}
		\frac{dR_s}{d\omega_{\Re}}=- \frac{(l+1/2)^2}{\omega^3_{\Re}\,\sqrt{\frac{(l+1/2)^2}{\omega^2_{\Re}}+a^2}}<0, 
	\end{eqnarray}
where the negative sign indicates that if the shadow radius increases, then the real part of QNMs decreases and vice versa. We can confirm it from Fig. 7 (right panel) where the slope of the function $R_s$ is negative for different values of $l$ numbers.  We can therefore write down the following relation
\begin{equation}
\frac{\partial T}{\partial r_{H}}=\frac{\partial T}{\partial R_s}\frac{\partial R_s}{\partial \omega_{\Re}} \frac{\partial \omega_{\Re}}{\partial r_{H}},
\end{equation}
where $r_H$ is the event horizon of the BH. In terms of the QNMs we can have thermodynamically stable BH provided
\begin{equation}\label{spheretwoa}
\frac{\partial T}{\partial R_{s}}>0\,\,\, \text{and}\,\,\, \frac{\partial \omega_{\Re}}{\partial r_{H}}<0,
\end{equation}
and thermodynamically unstable BH when
\begin{equation}\label{spheretwob}
\frac{\partial T}{\partial R_{s}}<0\,\,\, \text{and}\,\,\, \frac{\partial \omega_{\Re}}{\partial r_{H}}<0.
\end{equation}
In our case since we obtained $\partial T/\partial R_s<0$, we must have also $\partial \omega_{\Re}/\partial r_{H}<0$. We verify this fact in Fig. 7 (left panel), where the slope of the function $\omega_{\Re}$ with respect to $r_H$ is indeed negative for different $l$ numbers. However, the results will have more physical relevance if we study the stability using a more general case by varying the parameters $a$ and $b$, for polar and equatorial geodesics. The typical shadow radius for the polar modes given by Eq.~\eqref{R2} has a closed analytical form and it is much easier to study the stability. To have a more general picture, we need to study also the stability using the equatorial modes. For the case of equatorial modes, the expression for the typical shadow radius is more involved, and one is forced to perform numerical methods. Finally, it will be also interesting to compare both results and see to what extend they are complementary to each other. We plan to extend our approach here and study it in more details in the near future.

\section{Constraints on the quantum corrections from the S2 star orbit near the Sgr A$^{\star}$ BH}

One of the methods to study the nearby geometry of BH in the Milky Way galactic center is to analyze orbits of S cluster around Sgr A$^{\star}$ \cite{nucita,nucita1,alex}. In fact, it was argued that one can use the S cluster stars to set constraints on the BH mass (in the present work we shall assume $4.07\times 10^6 M_{\odot}$ \cite{Gillessen:2017}).  Moreover the motion of S2 star has been used to constrain different models for the dark matter distribution inside the inner galactic region, such as the dark matter spike model investigated in Ref. \cite{np}. In the present work, we are going to use the S2 orbit data collected during the last few decades (see, for more details \cite{Do:2019}), to fit the quantum deformed BH and see whether quantum corrections can mimic for example the effect of dark matter in the galactic center.

\begin{figure}
\includegraphics[scale=0.81]{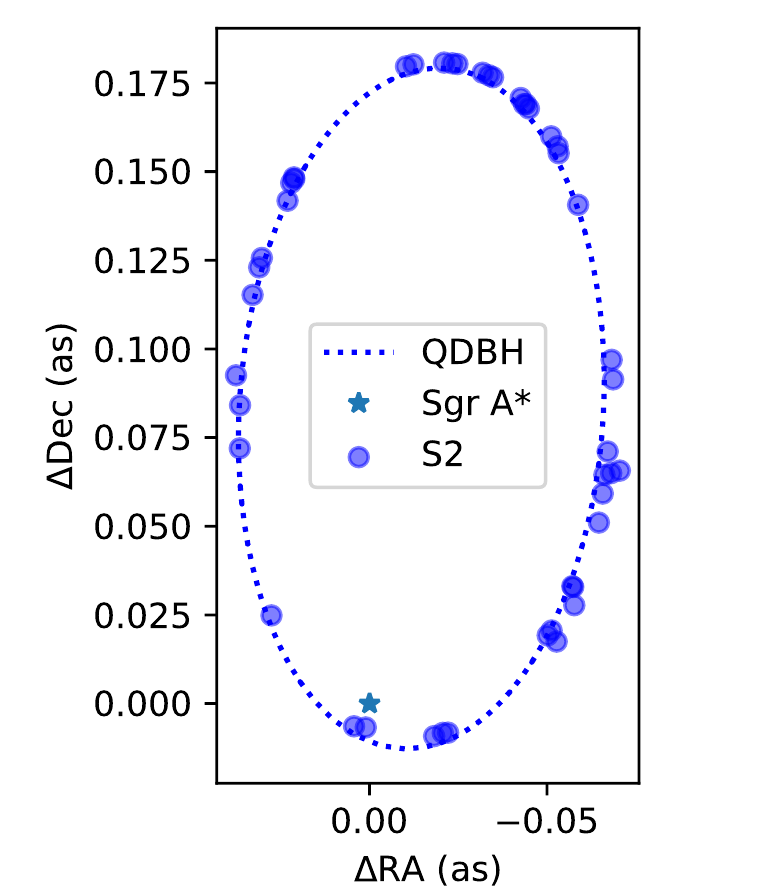}
\centering
    \caption{The plot shows the orbit of S2 star around the Sgr A$^\star$ BH as a model fitting using the quantum deformed BH. We used the observational data from \cite{Do:2019} and the best fit values obtained for deformed parameter $b$ and the black hole mass $M$.}
    \label{orbit}
\end{figure}

\begin{figure}
\centering
\includegraphics[width=3.6in]{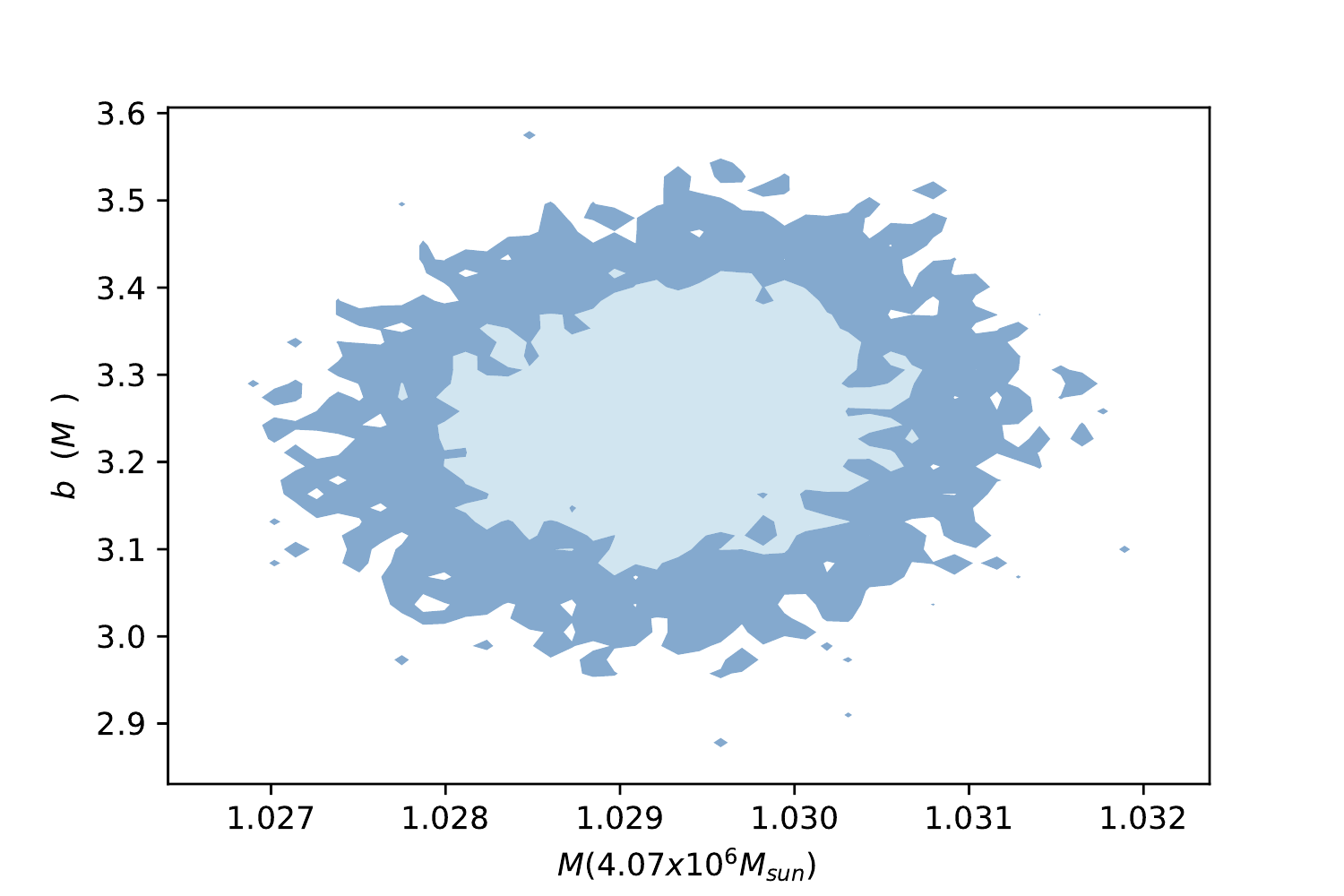}
    \caption{The plot shows the constrained quantum deformed parameter $b$ with 68\% and 96\% confidence contours.}
    \label{contour}
\end{figure}

For simplicity, we shall neglect the spinning BH parameter in this work. To fit our BH model, we have to solve the equations of motion numerically (see, for more details  \cite{rueda}) using few orbit parameters such as the inclination angle ($i$), the argument of periapsis  ($\omega$), the angle to the ascending node ($\Omega$), the semi-major axis ($a$) and finally the eccentricity ($e$) of the orbit. The best-fitting values for the parameter $b$ and the BH mass $M$ are derived from the MCMC analysis.

The observational data and the best-fitting orbit for the S2 are shown in Fig.~\ref{orbit}, where the star denotes the position of Sgr A$^\star$. We took the uniform priors in the interval $b \sim [0,10]$ $M$ and found the best fitting values for the $68\%$ confidence level of parameters  $b=3.24^{+0.11}_{-0.11} M$ while for the  $96\%$ confidence level we found $b=3.24^{+0.22}_{-0.22} M$ as shown in Fig. \ref{contour}. For the BH mass we obtain the best fitting values $M=1.029^{+0.001}_{-0.001}$ (in units $4.07\times 10^6 M_{\odot}$) within $68\%$ confidence and $M=1.029^{+0.002}_{-0.002}$ within $96\%$ confidence level, respectively.

	\section{Quasi-periodic oscillations\label{secqpos}}
	Introducing the relevant universal constants, the metric~\eqref{metric} takes the form
	\begin{align}
		&g_{t t}= -c^2 \Big[1-\frac{r^2[1-f(r)]}{\rho^2}\Big],\;g_{r r}= \frac{\rho^2}{r^2 f(r)+a^2}, \nonumber\\
		&g_{t \phi }=-\frac{car^2[1-f(r)]\sin^2\theta}{\rho^2},\;g_{\theta  \theta }=\rho^2=r^2+a^2 \cos ^2\theta ,\nonumber\\
		\label{metric2}&g_{\phi  \phi }=\Big[\frac{(r^2+a^2)^2-a^2[r^2 f(r)+a^2]\sin^2\theta}{\rho^2}\Big] \sin ^2\theta ,\\
		&f(r)=\sqrt{1-\frac{b^2}{r^2}}-\frac{2GM}{c^2r}=\sqrt{1-\frac{b_0^2}{x^2}}-\frac{2}{x}.\nonumber
	\end{align}
	From now on we introduce the dimensionless parameters ($a_0$, $b_0$, $x$) defined by
	\begin{equation}\label{dim}
		a_0\equiv\frac{a}{r_g},\quad b_0\equiv\frac{b}{r_g},\quad x\equiv\frac{r}{r_g},\quad r_g\equiv\frac{GM}{c^2}.
	\end{equation}
	Recall that for the Kerr BH $b_0=0$.
	
	For the numerical calculations to be carried out in this section, we take $M_\odot=1.9888\times 10^{30}$ (solar mass), $G=6.673\times 10^{-11}$ (gravitational constant), and $c=299792458$ (speed of light in vacuum) all given in SI units. These same constants will be written explicitly in some subsequent formulas of this section.
	
	\begin{figure}[h]
		\centering
		\includegraphics[width=0.49\textwidth]{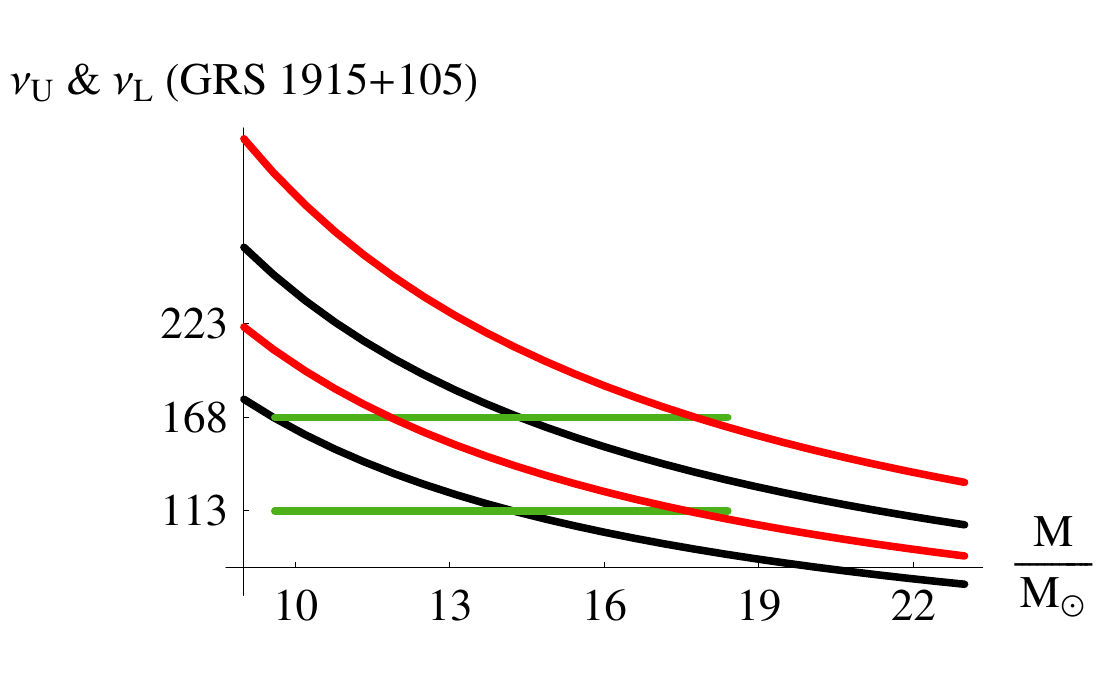}
		\caption{Fitting the particle oscillation upper and lower frequencies to the observed frequencies (in Hz) for the microquasar GRS 1915+105 at the 3/2 resonance radius. In the black plots the microquasar is treated as a quantum deformed Kerr BH given by~\eqref{metric2} with $b_0=0.6035$. The upper black curve represents $\nu_U=\nu_\theta$ and the lower black curve represents $\nu_L=\nu_r$ with $\nu_U/\nu_L=3/2$, and the green curve represent the mass error band as given in~\eqref{pr3}. The black curves cross the mass error bands well in the middle ensuring a good curve fitting. In the red plots the microquasar is treated as a Kerr BH $b_0=0$. The red plots also cross the mass error bands but at the rightmost points. A description of the astrophysical object by a Kerr BH does not provide a good curve fitting to justify the occurrence of the 3/2 resonance.}\label{FigGRS-I}
	\end{figure}
	\begin{figure}[h]
		\centering
		\includegraphics[width=0.49\textwidth]{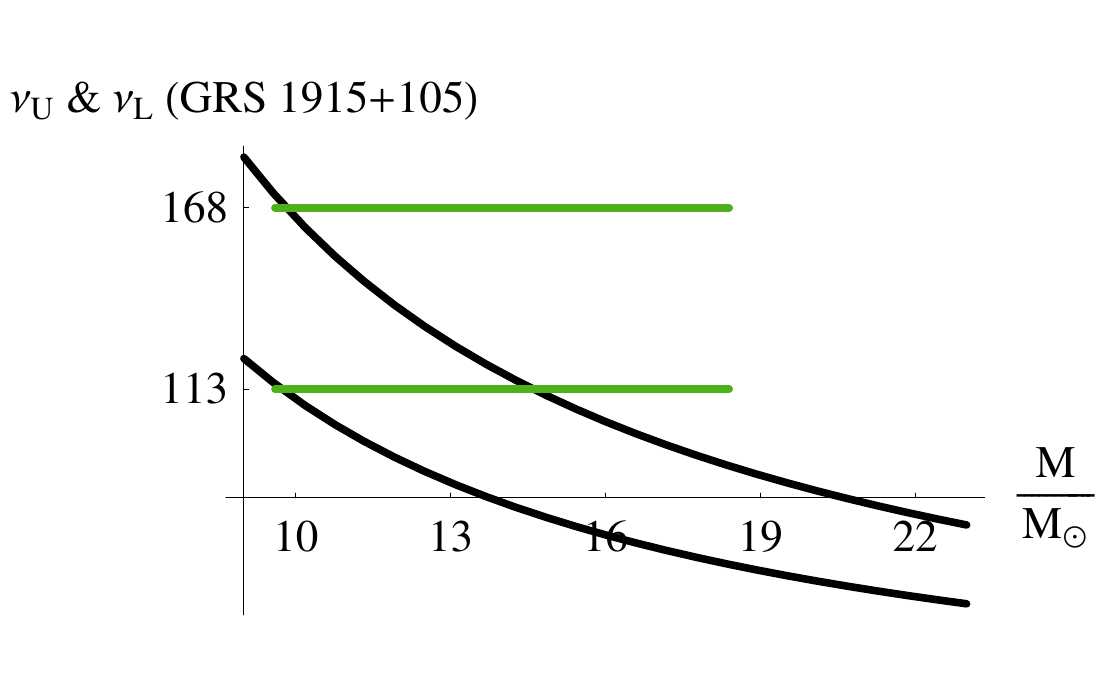}
		\caption{Fitting the particle oscillation upper and lower frequencies to the observed frequencies (in Hz) for the microquasar GRS 1915+105 at the 3/2 resonance radius. In the black plots the microquasar is treated as a quantum deformed Kerr BH given by~\eqref{metric2} with $b_0=1.2149$, which is an upper limit for $b_0$. The upper black curve represents $\nu_U=\nu_\theta$ and the lower black curve represents $\nu_L=\nu_r$ with $\nu_U/\nu_L=3/2$, and the green curve represent the mass error band as given in~\eqref{pr3}. The black curves cross the mass error bands at the leftmost points.}\label{FigGRS-U}
	\end{figure}
	\begin{figure}[h]
		\centering
		\includegraphics[width=0.49\textwidth]{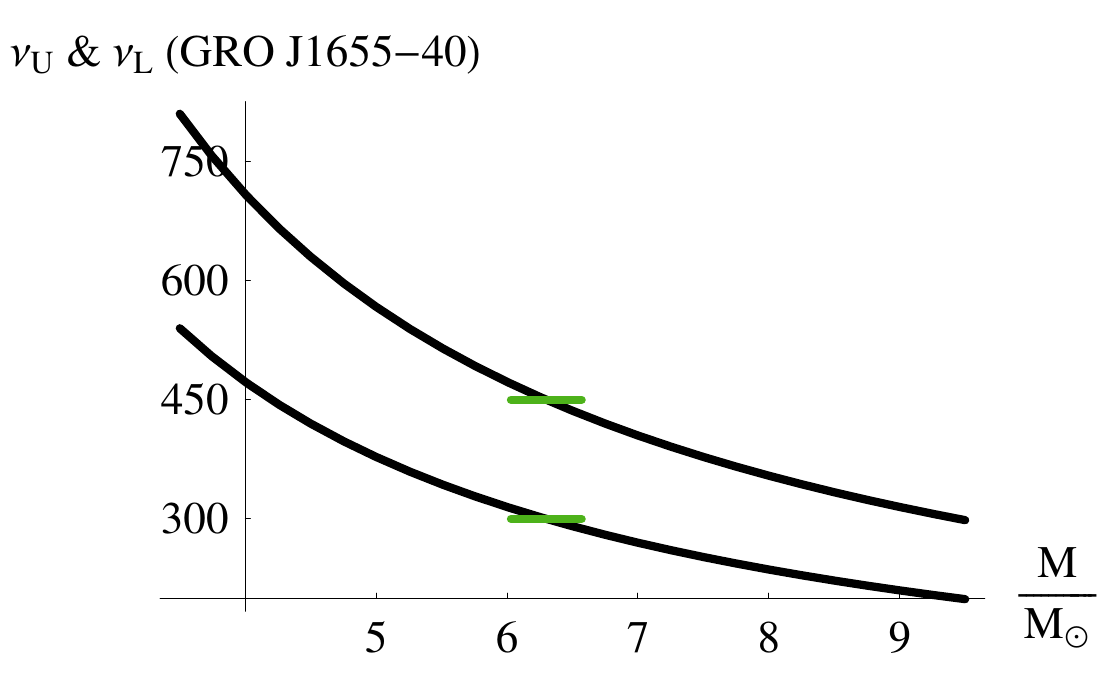}
		\caption{Fitting the particle oscillation upper and lower frequencies to the observed frequencies (in Hz) for the microquasar GRO J1655-40 at the 3/2 resonance radius. In the black plots the microquasar is treated as a quantum deformed Kerr BH given by~\eqref{metric2} with $b_0=1.0787\, {\rm i}$ (${\rm i}^2=-1$). The upper black curve represents $\nu_U=\nu_\theta$ and the lower black curve represents $\nu_L=\nu_r$ with $\nu_U/\nu_L=3/2$, and the green curve represent the mass error band as given in~\eqref{pr1}.}\label{FigGRO}
	\end{figure}
	\begin{figure}[h]
		\centering
		\includegraphics[width=0.49\textwidth]{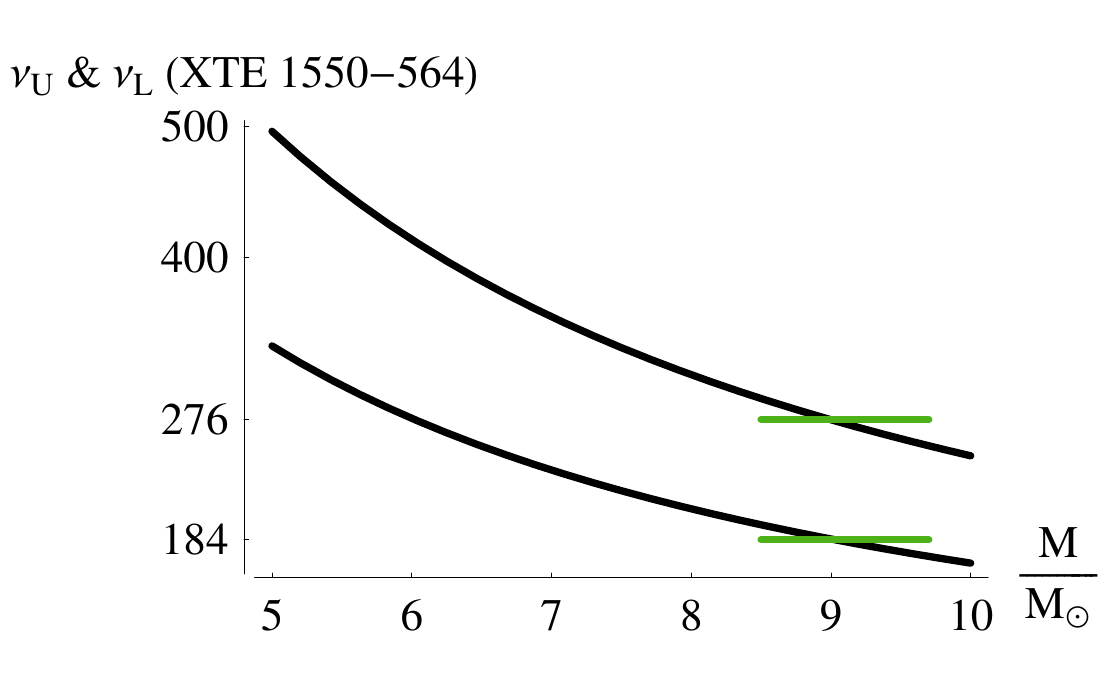}
		\caption{\centering Fitting the particle oscillation upper and lower frequencies to the observed frequencies (in Hz) for the microquasar XTE J1550-564 at the 3/2 resonance radius. In the black plots the microquasar is treated as a quantum deformed Kerr BH given by~\eqref{metric2} with $b_0=1.4825\, {\rm i}$ (${\rm i}^2=-1$). The upper black curve represents $\nu_U=\nu_\theta$ and the lower black curve represents $\nu_L=\nu_r$ with $\nu_U/\nu_L=3/2$, and the green curve represent the mass error band as given in~\eqref{pr2}.}\label{FigXTE}
	\end{figure}

	The power spectra of the galactic microquasar GRO J1655-40 reveal two peaks at 300 Hz and 450 Hz~\cite{res}, representing, respectively, the possible occurrence of the lower $\nu_L=300$ Hz QPO, and of the upper $\nu_U=450$ Hz QPO. Similar peaks have been obtained for the microquasars XTE J1550-564 and GRS 1915+105 obeying the remarkable relation, $\nu_U/\nu_L=3/2$~\cite{qpos1}. Some of the physical {quantities} of these three microquasars and their uncertainties are as follows~\cite{res,res2}:
	\begin{multline}\label{pr3}
		\text{GRS 1915+105 : }\frac{M}{M_\odot}=14.0\pm 4.4,\;a_0=0.99\pm 0.01\\
		\nu_U=168\pm 3 \text{ Hz},\;\nu_L=113\pm 5 \text{ Hz}.
	\end{multline}
	\begin{multline}\label{pr1}
		\text{GRO J1655-40 : }\frac{M}{M_\odot}=6.30\pm 0.27,\;a_0=0.70\pm 0.05\\
		\nu_U=450\pm 3 \text{ Hz},\;\nu_L=300\pm 5 \text{ Hz},
	\end{multline}
	\begin{multline}\label{pr2}
		\text{XTE J1550-564 : }\frac{M}{M_\odot}=9.1\pm 0.6,\;a_0=0.405\pm 0.115\\
		\nu_U=276\pm 3 \text{ Hz},\;\nu_L=184\pm 5 \text{ Hz},
	\end{multline}
	
	The twin values of the QPOs are most certainly due to the phenomenon of parametric resonance~\cite{res3,res4} where the in-falling charged particles perform radial and vertical oscillations around almost circular orbits with local radial and vertical frequencies denoted by ($\Omega_r,\,\Omega_\theta$), respectively. For the case of uncharged spinning BH ($\Omega_r,\,\Omega_\theta$) are given by~\cite{qposknb}
	\begin{align}\label{Omega}
		&\Omega_{r}^2\equiv (\partial_{r}\Gamma^{r}_{ij}-4\Gamma^{r}_{ik}\Gamma^{k}_{rj})u^iu^j,\qquad (i,\,j,\,k=t,\,\phi),\nonumber\\
		&\Omega_{\theta}^2\equiv (\partial_{\theta}\Gamma^{\theta}_{ij})u^iu^j,\qquad (i,\,j=t,\,\phi),
	\end{align}
	where the summations extend over ($t,\,\phi$). In obtaining these expressions we assumed that the main motion of the particle is circular in the equatorial plane ($\theta=\pi/2$) where the particle exhibits radial and vertical oscillations. The circular motion is stable only if $\Omega_r^2>0$ and $\Omega_\theta^2>0$. In the equatorial plane the four-velocity vector of the particle has only two nonzero components $u^{\mu}=(u^t,\,0,\,0,\,u^{\phi})=u^t(1,\,0,\,0,\,\omega)$, where $\omega=d \phi/d t$ is the angular velocity of the test particle. They are related by~\cite{qposknb}
	\begin{align}\label{velocity}
		&\omega =\frac{-\partial _r g_{t \phi }\pm \sqrt{\left(\partial _r g_{t \phi }\right)^2-\partial _r g_{t t} \partial _r g_{\phi  \phi }}}{\partial _r g_{\phi
				\phi }}=\Omega_\phi,\nonumber\\
		&u^t=\frac{c}{\sqrt{-\left(g_{t t}+2 \partial _r g_{t \phi } \omega +g_{\phi  \phi } \omega ^2\right)}},\nonumber\\
		&u^{\phi }=\omega  u^t .
	\end{align}
	It is understood that all the functions appearing in~\eqref{Omega} and~\eqref{velocity} are evaluated at $\theta=\pi/2$. Here $\Omega_{\phi}$ is the same entity defined in~\eqref{Omegaf}.
	
	The locally measured frequencies ($\Omega_{r},\,\Omega_{\theta}$) are related to the spatially-remote observer's frequencies ($\nu_{r},\,\nu_{\theta}$) by
	\begin{align}\label{pr4}
		&\nu_r=\frac{1}{2\pi}~\frac{1}{u^t}~\Omega_{r},
		&\nu_\theta=\frac{1}{2\pi}~\frac{1}{u^t}~\Omega_{\theta}.
	\end{align}
	Using the form~\eqref{metric2} of the metric one expresses ($\nu_{r},\,\nu_{\theta}$) in terms of the dimensionless parameters ($x$, $a_0$, $b_0$) and ($c$, $M$, $G$). These expressions are sizable to be given here and we content ourselves to plot them versus $M$ for ($x$, $a_0$, $b_0$) held fixed. Perfect curve fitting of the particle oscillation upper and lower frequencies to the observed frequencies (in Hz) for the microquasar GRS 1915+105~\eqref{pr3} are shown in Fig.~\ref{FigGRS-I}. In the black plots the microquasar is treated as a quantum deformed Kerr BH given by~\eqref{metric2} and taking the coordinates of $(b_0,\,x)$ to be the values where the 3/2 resonance occurs ($\nu_U=\nu_\theta=168$ Hz and $\nu_L=\nu_r= 113$ Hz yielding $b_0=0.6035$ and $x=5.0076$). In the red plots the microquasar is treated as a Kerr BH ($b_0=0$). We see from this figure that the black plots cross the mass error bands exactly in the middle point. The red plots also cross the mass error bands but almost at the rightmost points. In Fig.~\ref{FigGRS-U} we provide the upper limit of $b_0$ that provides a mediocre curve fitting of the particle oscillation upper and lower frequencies to the observed frequencies: $b_{0\,\text{max}}=1.2149$.
	
	Based on the previous analysis, we restrict the values of $b_0$ to lie between the smallest and largest values we obtained above:
	\begin{equation}\label{bz}
		0\leq b_0\leq 1.2149 .
	\end{equation}
	
	Taylor expansions of ($\nu_{r},\,\nu_{\theta}$)~\eqref{pr4} about $b_0=0$ introduce corrections to the corresponding Kerr expressions
	\begin{widetext}
		\begin{align}\label{nrt}
			&\nu _r=\nu _{r\,\text{Kerr}}-\frac{c^3 [a_0^3-4 a_0^2 (1-x) \sqrt{x}+a_0 (3-11 x) x+9 x^{5/2}]}{8 \pi  G M x^2 \left(a_0+x^{3/2}\right)^2 \sqrt{8 a_0 \sqrt{x}-3
					a_0^2-(6-x) x}} b_0^2+\mathcal{O}(b_0^4),\\
			&\nu _{\theta }=\nu _{\theta\,\text{Kerr}}+\frac{c^3 (2 a_0^2 x^{3/2}-a_0^3-4 a_0 x^2+x^{7/2})}{8 \pi  G M x^2 (a_0+x^{3/2})^2 \sqrt{3 a_0^2-4 a_0 \sqrt{x}+x^2}} b_0^2+\mathcal{O}(b_0^4),
		\end{align}
	\end{widetext}
    where ($\nu _{r\,\text{Kerr}},\,\nu _{\theta\,\text{Kerr}}$) are given in~\cite{Kerr1,Kerr2}.
	
	For the other two microquasars~\eqref{pr1} and~\eqref{pr2} it is not possible to have a curve fitting of the particle oscillation upper and lower frequencies to the observed frequencies if $b_0^2>0$. If we allow $b^2$ to be negative as we mentioned in Sec.~\ref{secrqd}, assuming $B_0^2\equiv -b_0^2>0$, we obtain the following limits of $B_0$ for the microquasars GRO J1655-40 and XTE J1550-564, respectively:
	\begin{align}
		& \text{GRO J1655-40: }\qquad 1.0475\lesssim B_0\lesssim 1.1043,\\
		& \text{XTE J1550-564: }\qquad 1.4424\lesssim B_0\lesssim 1.5146.
	\end{align}
	The corresponding curve fittings are presented in Figs.~\ref{FigGRO} and~\ref{FigXTE}.

	\section{Conclusion}\label{conc}
	In this paper, we studied the relationship between QNMs and BH shadow for the quantum deformed Kerr spacetime. By using the geometric-optics correspondence between the parameters of a QNMs and the conserved quantities along geodesics, in the eikonal limit, the real part of QNMs is shown to be related with the Keplerian frequency in the equatorial plane orbits. Further, new formulas for the shadow radius having a viewing angle $\theta_0=\pi/2$ and $\theta_0=0$ are derived. As a special case, we were able to derive the equation for the typical shadow radius reported earlier in the literature as well as the static limit.  We considered the quantum deformed BH as a seed solution in the NJAA to obtain a spinning quantum deformed BH and studied its shadow. The typical size for the shadow radius was found to be increased with the increase of the parameter $b$. We analyzed the equatorial modes and polar modes in addition to the QPOs to constrain the quantum deforming parameter. Importantly, we have explored the thermodynamical stability of the BH using the properties of its shadow. Moreover the thermodynamical stability is also studied by the properties of the QNMs frequencies using its inverse relation with the shadow radius. As an example we considered the shadow radius for the viewing angle $\theta_0=0$ (and alternatively $\theta_0=\pi)$  to obtain new inequalities in terms of the QNMs and shadow radius.
	
	In addition to that, we have also explored the properties of the quantum deformed BH in astrophysical problems, such as the orbit fitting of the motion of the S2 star in our galactic center. We have constraint the quantum-deforming parameter $b_0$ and found the best fit to be $b_0 \sim 3.24$.
	
	The investigation of QPOs allows us to enlarge the domain of the quantum-deforming parameter $b_0^2$ to include negative values. Curve fitting to the observed values restricted $\sqrt{|b_0^2|}$ to be of order unity. We thus found that the upper bound of the parameter $b_0$ found by the QPOs is out of factor 3 compared to the best fit obtained by the orbit of S2. This is possibly linked to the fact that inside the region between the S2 and the BH there is additional contribution of matter, namely dark matter. Thus we can exclude the possibility that quantum deformed BH metric can mimic at the same time the dark matter at the galactic center and the spacetime region of Galactic microquasars considered in this work. 
	
	\acknowledgements
 The work of QW and MJ is supported in part by the National Key Research and Development Program of China Grant No.2020YFC2201503, the Zhejiang Provincial Natural Science Foundation of China under Grant No. LR21A050001, the Zhejiang Provincial Natural Science Foundation of China under Grant No.LY20A050002, the Fundamental Research Funds for the Provincial Universities of Zhejiang in China under Grant No. RF-A2019015, and National Natural Science Foundation of China under Grant No. 11675143.

\end{document}